\documentclass[preprint]{aastex}

\usepackage[utf8x]{inputenc}
\usepackage[english]{babel}
\usepackage{fontenc}
\usepackage{graphicx}

\usepackage[breaklinks,dvipdfm]{hyperref}

\usepackage{amsmath}
\usepackage{amsfonts}
\usepackage{amssymb}
\usepackage{mathrsfs}

\usepackage{multicol}

\DeclareFontFamily{U}{mathb}{\hyphenchar\font45}
\DeclareFontShape{U}{mathb}{m}{n}{
      <5> <6> <7> <8> <9> <10> gen * mathb
      <10.95> mathb10 <12> <14.4> <17.28> <20.74> <24.88> mathb12
      }{}
\DeclareSymbolFont{mathb}{U}{mathb}{m}{n}
\DeclareFontSubstitution{U}{mathb}{m}{n}
\DeclareMathSymbol{\leftmoon}      {0}{mathb}{"4B}

\usepackage[squaren,Gray]{SIunits}
\addunit{\erg}{erg}
\addunit{\km}{km}

\usepackage{color}

\usepackage{soul}
\usepackage{empheq}

\shorttitle{Lunar accretion from a fluid disk}
\shortauthors{Salmon \& Canup}

\begin{document}

\title{Lunar accretion from a Roche-interior fluid disk}

\author{Julien~Salmon\footnote{Corresponding author.}}
%\affil{Southwest Research Institute}
%\affil{Department of Space Studies\\1050 Walnut Street, Suite 300, Boulder, CO 80302, USA}
\email{julien@boulder.swri.edu}
\and
\author{Robin~M.~Canup}
\email{robin@boulder.swri.edu}
\affil{Southwest Research Institute}
\affil{Department of Space Studies\\1050 Walnut Street, Suite 300, Boulder, CO 80302, USA}
\author{}
\author{Accepted for publication in the Astrophysical Journal on 02 October 2012}

\begin{abstract}
We use a hybrid numerical approach to simulate the formation of the Moon from an impact-generated disk, consisting of a fluid model for the disk inside the Roche limit and an N-body code to describe accretion outside the Roche limit. As the inner disk spreads due to a thermally regulated viscosity, material is delivered across the Roche limit and accretes into moonlets that are added to the N-body simulation. Contrary to an accretion timescale of a few months obtained with prior pure N-body codes, here the final stage of the Moon's growth is controlled by the slow spreading of the inner disk, resulting in a total lunar accretion timescale of $\sim \unit{10^2}{years}$. It has been proposed that the inner disk may compositionally equilibrate with the Earth through diffusive mixing, which offers a potential explanation for the identical oxygen isotope compositions of the Earth and Moon. However, the mass fraction of the final Moon that is derived from the inner disk is limited by resonant torques between the disk and exterior growing moons. For initial disks containing $< 2.5$ lunar masses ($M_\leftmoon$), we find that a final Moon with mass $> 0.8 M_\leftmoon$ contains $\le 60\%$ material derived from the inner disk, with this material preferentially delivered to the Moon at the end of its accretion.   
\end{abstract}


\maketitle

\section{Introduction}
The generally accepted scenario for the formation of the Moon involves the oblique impact of a roughly Mars-sized object with the proto-Earth \citep{hartmann75,cameron76,stevenson87,canup04b}. Numerical simulations of such an impact, using primarily Smoothed Particle Hydrodynamics (SPH) methods, have shown that the impactor is destroyed (either during the impact, or via post-impact tidal disruption), and that a circumterrestrial disk is formed that contains up to $\sim 2M_\leftmoon$ of iron-depleted material \citep{benz86, benz87, benz89, cameron91, cameron97, canup01, canup04a, canup08}.  The silicate disk is initially a mixture of vapor and melt, containing $\sim O(10\%)$ vapor by mass \citep{canup04a}. Typically about $\sim 20$ to $50\%$ of the disk material is predicted to have initial orbits exterior to the Roche limit for lunar density material, $a_R$, with $a_R \approx 2.9R_{\oplus}$ where $R_{\oplus}$ is the Earth's radius.

Prior works have used direct N-body simulations to model the accumulation of the Moon from such an impact-generated disk, describing the disk with $N = 10^3$ to $10^4$ particles that are each of order $\unit{10^2}{\km}$ in radius \citep{ida97,kokubo00}. The N-body models depict an extremely rapid disk evolution, with material interior to the Roche limit spreading outward on a timescale of order 1 month.  Typically a single massive moon accretes in a year or less at an average distance of  $\langle a\rangle \approx 1.3a_R$ \citep{ida97,kokubo00}. That a particulate protolunar disk would spread so rapidly was anticipated by earlier estimates of \citet{ward78}, who pointed out that a disk of particles containing sufficient mass to produce the Moon would be prone to gravitational instability and local clumping. Exterior to the Roche limit, instability-produced clumps would form permanent aggregates and seed the growth of the Moon. But interior to the Roche limit, such clumps are continually sheared apart by planetary tidal forces, and this process generates a large viscosity that drives a $\sim$ lunar mass Roche-interior disk of particles to spread in less than a year \citep{ward78, takeda01} (see Section \ref{subsubsection_instability_vicosity}. Both processes can be seen directly in the N-body simulations \citep{kokubo00}.

N-body protolunar disk models assume a disk of condensed particles, neglecting the presence and creation of vapor as the disk evolves. This is probably a reasonable approximation for material orbiting outside the Roche limit. Immediately after the impact, disk material is primarily in a condensed state \citep{canup04a}. Outside the Roche limit, collisions between orbiting particles lead to accretional growth. A rough estimate of the heat liberated by accreting the Moon is its gravitational binding energy, which implies an energy released per unit mass of the Moon of $E_b \sim (3/5)GM_\leftmoon/R_\leftmoon \sim \unit{2\times10^{10}}{\erg\usk\reciprocal\gram}$, where $G=\unit{6.67\times 10^{-8}}{\centi\cubicmetre\usk\reciprocal\gram\usk\rpsquare\second}$ is the gravitational constant, and $M_\leftmoon = \unit{7.35 \times 10^{25}}{\gram}$ and $R_\leftmoon = \unit{1738}{\km}$ are the Moon's mass and radius.  Even in the limit that all of this energy is retained by the Moon, the expected extent of vapor production as the Moon accretes is small because $E_b$ is much smaller than the latent heat of vaporization of silicate, $l_v \approx \unit{2\times10^{11}}{\erg\usk\reciprocal\gram}$.

However, there is an inherent inconsistency in describing the Roche-interior disk with an N-body particulate model. An approximately lunar mass disk of condensates (solid or melt) interior to the Roche limit will be subject to the instability-induced viscosity described above. Such a disk spreads so rapidly and the viscously generated heat is so large that the disk would likely substantially vaporize as it evolves \citep{thompson88}, invalidating the model's assumption of a particulate disk. A vapor disk would be gravitationally stable, and therefore not be subject to the instability-induced viscosity seen in the N-body simulations. \citet{thompson88} were the first to recognize this important point, and proposed that the protolunar disk would instead evolve in a two-phase, vertically mixed vapor-melt state, with a viscous dissipation rate regulated by the rate at which a $\sim \unit{2000}{\kelvin}$ silicate vapor photosphere could cool. This thermally regulated viscosity implies a much longer disk spreading timescale of $\sim \unit{10^2}{years}$ \citep{thompson88} (see Section \ref{subsubsection_thermal_viscosity}).

Recently \citet{ward12} has derived an analytical description for the vertical structure of a two-phase protolunar disk inside the Roche limit, including both the vertically well-mixed case explored by \citet{thompson88}, in which the vapor mass fraction is very low, and a new alternative class of solutions in which the vapor initially contains the majority of the disk's initial mass. The latter implies a stratified disk structure, in which a portion of the disk's mass settles to the mid-plane as melt and undergoes rapid viscous spreading, while the remainder of the disk is contained in a gravitationally stable vapor atmosphere. The vapor component of the disk requires $\sim \unit{10^2}{years}$ to deplete itself through condensation, and material is ultimately supplied to the Roche exterior region over this timescale. Thus both the well-mixed and stratified disk models imply a similarly protracted timescale for the inner disk's overall evolution that is $\sim \unit{10^2}{years}$.

We here develop a new lunar accretion model that describes the Roche-interior region as a fluid disk, while material outside the Roche limit is tracked using direct N-body simulation. The inner disk evolves viscously and interacts with outer bodies through resonant torques at the strongest Lindblad resonances. Material from the inner disk that viscously spreads beyond the Roche limit accretes to form new moonlets that are added to the N-body simulation, while inner disk material spreading onto the Earth is removed from the disk. This hybrid construct allows us to model a slowly evolving inner disk that spreads in $\sim \unit{10^2}{years}$ as suggested by thermodynamical models \citep{thompson88,ward12}, while also directly simulating the rapid accretion expected among condensed material orbiting outside the Roche limit.  

Our overall objective is a more physically motivated model of the Moon's accretion, and improved estimates of its formation timescale and initial orbital position.  These quantities are related to several outstanding issues, including the Moon's initial thermal state, the potential for chemical equilibration between the protolunar disk and the Earth prior to the Moon's accumulation \citep{pahlevan07}, and the initial disk mass and angular momentum required to produce a lunar mass Moon. In section 2 we describe in detail our numerical model. In section 3 we use our code to reproduce results from pure N-body simulations by \citet{ida97}. In section 4 we perform hybrid model simulations with an inner fluid disk, and study the influence of the disk's initial parameters. Results are then discussed in section 5.

\section{The model}
Our numerical model is built around the symplectic integrator SyMBA \citep{duncan98}. We have paired it with a simple analytical model for a Roche-interior fluid disk that evolves under the influence of viscous spreading and resonant torques due to interactions with orbiting objects at their $0^{th}$ order Lindblad resonances. The inner disk mass decreases as material spreads onto the planet or outward beyond the Roche limit; mass spreading beyond the Roche limit is assumed to accrete into new moonlets that are then added to the N-body code. We assume that the inner disk has a uniform surface density $\sigma = \sigma(t)$ and viscosity $\nu = \nu(t)$ with radius. The inner disk's evolution is computed by estimating the rate of change of its edges due to the viscous and resonant torques. These simplified inner disk treatments are detailed in Appendix \ref{appendix_disk_mass} and \ref{appendix_resonances}. 

\subsection{Viscosity model}
We characterize the inner disk by a single, time-dependent viscosity that is a function of the disk's surface density $\sigma$. We envision a silicate disk that is initially two-phase (vapor/melt), and we assume that both components co-evolve.  We adopt the argument of \citet{thompson88} that the inner disk's viscosity will be limited by the rate at which a $\sim \unit{2000}{\kelvin}$ disk photosphere can radiatively cool, so long as there is vapor present.  Once the disk mass and the associated rate of viscous dissipation is low enough that all of the vapor can condense, we assume that the inner disk's viscosity will be comparable to that of a purely condensate disk subject to local gravitational instabilities \citep{ward78}.  

\subsubsection{Instability induced viscosity}\label{subsubsection_instability_vicosity}
For a $\sim$ lunar mass, Roche-interior disk composed of melt or solids, local patch instabilities strongly increase the collision rate among disk particles through the formation of clumps that are continuously sheared apart by planetary tides \citep{ward78}. By introducing coherent particle motions, instabilities modify the transport of angular momentum in the disk and produce a characteristic viscosity \citep{ward78}
\begin{equation}
\nu_{WC} \sim \frac{\pi^2G^2\sigma^2}{\Omega^3},
\label{equ_viscosity_WC}
\end{equation}
where $\sigma$ is the disk surface density, and $\Omega$ is orbital frequency. This process and the resulting rate of angular momentum transport are observed in N-body numerical simulations of the protolunar disk \citep{takeda01} and dense planetary rings \citep{salo95, daisaka99, daisaka01}. The associated disk spreading timescale for a disk of radial scale $r$, $r^2/\nu$, is then
\begin{equation}
\tau_{WC} = \frac{r^2 \Omega^3}{\pi^2 G^2 \sigma^2} \sim 0.8 \left(\frac{M_d}{M_{\leftmoon}}\right)^{-2}\left(\frac{r_d}{a_R}\right)^{3/2}~\text{years}.
\label{tau_WC}
\end{equation}

The dissipation rate per unit area is $\dot{E_{\nu}} = (9/4)\sigma\nu_{WC}{\Omega}^2$ ($\dot{X}$ denotes the time derivative of $X$), implying a total energy per area dissipated as the disk spreads for a time ${\tau}_{WC}$ of $\dot{E_{\nu}}\tau_{WC}\sim (9/4)\sigma{r\Omega}^2$. For a uniform surface density disk, this represents a total liberated energy per unit disk mass of $\sim (9/4)(r{\Omega})^2 \sim 5\times10^{11} (a_R/r_d)~{\erg\usk\reciprocal\gram}$.  

\subsubsection{Thermally regulated viscosity}\label{subsubsection_thermal_viscosity}
The estimated energy released as an $\sim$ lunar mass disk spreads is thus comparable to the latent heat of vaporization for silicate ($l_v \sim \unit{2\times10^{11}}{\erg\usk\reciprocal\gram}$). If the inner disk spreads in $< \unit{1}{year}$ as implied by equation (\ref{tau_WC}), it is probable that it will retain any viscously dissipated heat, because the timescale for the disk to radiatively cool from its surfaces is much longer, of order decades \citep{thompson88, pritchard00}. This implies that a substantial fraction of a condensate disk would vaporize as it spreads due to an instability-induced viscosity. 

While a condensate disk would be subject to instabilities, a typical silicate vapor disk is gravitationally stable. Indeed, its Toomre parameter is \citep{toomre64}
\begin{equation}
Q = \frac{c_s\Omega}{\pi G \sigma} \approx 52\left(\frac{r_d}{R_{\Earth}}\right)^{-3/2}\left(\frac{T}{\unit{2000}{\kelvin}}\right)^{1/2}\left(\frac{\sigma}{\unit{10^7}{\gram\usk\centi\rpsquare\metre}}\right)^{-1},
\end{equation}
where $T$ is the disk's temperature and $c_s=\sqrt{\gamma R T/\mu}$, $\gamma=1.4$ is the adiabatic index, $R$ is the universal gas constant and $\mu=\unit{30}{\gram\usk\reciprocal\mole}$ is the molecular weight. Thus, a protolunar disk composed of silicate vapor with a photosphere temperature near the condensation point $(T\sim\unit{2000}{\kelvin})$ would have $Q > 1$, and would not be subject to the instability-induced viscosity. 

In the absence of strong dissipation, a vapor disk could cool and condense, with gravitational instabilities then re-developing in the condensed phase.  This in turn would heat the disk through increased dissipation. This feedback suggests that the dissipation rate in the inner disk will be limited by the rate at which the disk can radiatively cool from its surfaces \citep{thompson88}, with 
\begin{equation}
\frac{9}{4}\sigma\nu{\Omega}^2 = 2\sigma_{SB}T_p^4,
\end{equation}
and an associated viscosity \citep{thompson88}
\begin{equation}
\nu_{TS} \approx \frac{\sigma_{SB}T_p^4}{\sigma\Omega^2},
\label{equ_viscosity_TS}
\end{equation}
where $\sigma_{SB}$ is the Stefan-Boltzmann constant and $T_p$ is the disk's photospheric temperature. The corresponding spreading timescale is 
\begin{equation}
\tau_{TS} \approx 50 \left(\frac{r_d}{a_R}\right)^{-3} \left(\frac{T_p}{\unit{2000}{\kelvin}}\right)^{-4} \left(\frac{M_d}{M_{\leftmoon}}\right)~\text{years}.
\end{equation}

\subsubsection{Model used}{\label{section_model_used}}
Following the \cite{thompson88} disk model in which the liquid and vapor phases remain vertically well-mixed, our model assumes that both the vapor and condensed phases viscously evolve as a single unit. At each time step in our simulation, we compute both the instability-induced viscosity $\nu_{WC}$ and the radiation-limited viscosity $\nu_{TS}$ (with $T_p = \unit{2000}{\kelvin}$) at $r=r_d$.  If $\nu_{WC} > \nu_{TS}$, we assume that the disk self-regulates to a radiation-limited viscosity and set $\nu = \nu_{TS}$.  If $\nu_{WC} < \nu_{TS}$, the disk can lose energy via radiative cooling at a faster rate than it is generated by instabilities. At this point we assume the disk would condense, and therefore set $\nu = \nu_{WC}$. 

The ratio between the two viscosities is 
\begin{equation}
 \frac{\nu_{TS}}{\nu_{WC}} \sim \frac{\sigma_{SB} T_p^4\Omega}{\sigma^3\pi^2G^2} 
 \sim 5\times 10^{-3}\left(\frac{T_p}{\unit{2000}{\kelvin}} \right)^4 \left( \frac{r_d}{a_R}\right)^{-3/2} \left( \frac{\sigma}{\unit{10^7}{\gram\usk\centi\rpsquare\metre}}\right)^{-3}.
\end{equation}
For reference, a uniform one lunar mass disk extending from $R_{\oplus}$ to $a_R$ has a surface density of $\sim \unit{8\times10^6}{\gram\usk\centi\rpsquare\metre}$.
Our initial disks have $\nu_{TS}/\nu_{WC} < 1$ and evolve with a radiation-limited viscosity. As the disk spreads and loses mass, its surface density decreases, and since $(\nu_{TS}/\nu_{WC}) \propto \sigma^{-3}$, at some point $\nu_{WC} \sim \nu_{TS}$.  This transition occurs for 
\begin{equation}
\sigma_{trans} \sim \left(\frac{\sigma_{SB}T_p^4\Omega}{\pi^2G^2}\right)^{1/3} 
               \sim 1.7\times10^6 \left(\frac{T_p}{\unit{2000}{\kelvin}}\right)^{4/3}\left(\frac{r}{a_R}\right)^{-1/2}{\gram\usk\centi\rpsquare\metre}.
\end{equation}
This is equivalent to an inner disk mass of $M_d=\unit{1.6\times10^{25}}{\gram} \approx 0.2~M_{\leftmoon}$ for a uniform surface density between $r=R_{\Earth}$ and $r_d=a_R$. 

We assume that the inner disk maintains a uniform surface density with radius as it viscously expands.  Viscous expansion leads to mass transfer from the disk onto the planet, and to the outward expansion of the disk's outer edge (see Appendix \ref{appendix_disk_mass} for details).

\subsection{Spawning of moonlets}{\label{section_spawning}}
As disk material viscously spreads outward beyond the Roche limit, accretion becomes increasingly probable and the continuous nature of the disk is disrupted as discrete large objects form \cite[e.g.][]{kokubo00}. Our model approximates this transition by removing mass from the inner disk and adding new moonlets to the N-body simulation once the disk's outer edge expands beyond the Roche limit. Once $r_d \ge a_R$, we compute the characteristic fragment mass that would form from local gravitational instability, and assume that since it is at or beyond the Roche limit, it will be stable and not be tidally disrupted. This mass, and its corresponding angular momentum, is removed from the inner disk and added to the N-body code as a new discrete particle.

The mass $m_f$ of the fragment that would form from instabilities is \citep{goldreich73}
\begin{equation}
 m_f=\frac{16\pi^4\xi^2\sigma^3r_d^6}{M_\Earth^2},
\label{equ_fragment_mass}
\end{equation}
where $\xi$ is on the order of, but less than, unity. We set $\xi=0.3$. If the disk's outer edge is at the Roche limit $\left(r_d=2.9R_\Earth\right)$, then inner disks containing $1.5$ and $0.01 M_\leftmoon$ will form fragments of $\approx 3\times 10^{-3} M_\leftmoon$ and $10^{-9} M_\leftmoon$, respectively. This is comparable to the aggregate mass seen in the direct ``rubble pile'' N-body simulations of \citet{kokubo00}. To improve computational efficiency, we set the minimal mass of spawned fragments to $\approx 10^{-5} M_\leftmoon$. Smaller fragments would be formed by an inner disk containing $< 0.3 M_\leftmoon$. As we will later see, this only affects the very last stages of a given simulation, so we expect it to be of little influence on the outcome of a given simulation. We test the influence of this parameter in Section \ref{section_model_limitations}.

At each time step, we check whether the disk's outer edge lies beyond the Roche limit.  When this occurs, we compute the mass of a spawned fragment per the equation above as a function of $\sigma$. The mass of the spawned moonlet is removed from the inner disk. To conserve angular momentum, we first set the new body's semi-major axis to $r_d$, and then we compute the new disk's outer edge $r_d'$ so that $L_d' + L_f - L_d=0$, where $L_d$ and $L_d'$ are the inner disk's angular momentum before and after adding the new body, and $L_f$ is the added fragment's angular momentum. Additional details are in Appendix \ref{appendix_spawning}.

\subsection{Disk-satellite interactions}
We include resonant interactions between the disk and the growing moonlets, which lead to a positive torque on the exterior moonlets (whose orbits expand), and a negative torque on the inner disk (whose outer edge contracts). Such interactions are important because, e.g., sufficiently massive moonlets can initially confine the disk edge within the Roche limit and delay the spawning of additional moonlets \citep{charnoz10}.    

As a first approximation, we consider only the strongest $0^{th}$ order inner Lindblad resonances, in which the ratio of the mean motion at a location in the disk to that of an exterior moonlet is a ratio of integers with $\left(m:m-1\right)$. To compute the resonant torque, we use the formalism of \citet{goldreich80}. 

The total torque $T_s$ exerted by the inner disk on an exterior satellite per unit satellite mass is found by summing the torques due to all the $0^{th}$ order resonances that fall in the disk (see Appendix \ref{appendix_resonances}),

\begin{equation}
\frac{T_s}{M_s} = \left(\frac{\pi^2}{3}\mu_sG\sigma a_s\right)C(m),
\label{equ_disktorque_on_sat}
\end{equation}
where $M_s$ is the satellite's mass, $\mu_s = M_s/M_\Earth$, $C(m) = \displaystyle{\sum_{m=2}^{m_*}2.55m^2(1-1/m)}$ and $m_*$ is the highest $m$ for which resonance $(m:m-1)$ falls in the disk. When $m_* < 2$, the satellite is far enough from the disk that its 2:1 resonance (which is the most distant $0^{th}$ order resonance) is no longer in the disk, and in this case $T_s = 0$.  For a satellite orbiting close to the disk, we impose an upper limit on $m_*$ by considering only those resonances that are radially separated from the satellite's orbit by a distance greater than the satellite's Hill radius ($R_H = a_s(M_s/3M_\oplus)^{1/3}$), or those resonances for which $\left(1-1/m\right)^{2/3} \le 1 - \left(M_s/3M_\oplus\right)^{1/3}$. This excludes from the torque calculation the approximate region immediately surrounding the satellite's semi-major axis within which particles undergo horseshoe orbits.

To compute the resulting orbital evolution of the satellite we adopt the approach of \citet{papaloizou00}. We define an orbital migration timescale, $t_m$, due to the torque on the satellite associated with all of its $0^{th}$ order resonances that fall in the disk, $t_m=L_s/T_s$, where $L_s$ is the satellite's orbital angular momentum. We then apply an additional acceleration to the satellite, given by $\mathbf{a}_{mig} = (\mathbf{v}/{t_m})$, where $\mathbf{v}$ is the satellite's velocity. We include this as an additional ``kick'' of duration $\tau/2$ (where $\tau$ is the timestep) at the beginning and end of each step in the N-body code.

The total torque on the disk due to $N$ orbiting moonlets is $\displaystyle{T_d = {- \sum_{s=1}^{N}T_s}}$. We assume that moonlet torques cause a change in the disk's outer edge $r_d$, with $\dot{r}_d|_{moon} < 0$  (see Appendix \ref{appendix_resonances}) because external moons remove angular momentum from the disk.

\subsection{Model for Roche exterior particulate disk}
Beyond the Roche limit, we model the protolunar disk material by a collection of individual particles, with an initial power-law size distribution $N(m)dm \propto m^{-p}dm$, where $N(m)$ is the number of particles with a mass between $m$ and $m+dm$. In this section we describe our treatments of collisions between particles, and the tidal disruption of moonlets scattered close to the Earth.

\subsubsection{Tidal accretion criteria}
In the default version of SyMBA, all collisions result in inelastic mergers. However this is too simplified for objects orbiting near the Roche limit. We modified the code to include tidal accretion criteria \citep{ohtsuki93,canup95}, which near the Roche limit are a function of the impact energy, the mass ratio of the colliding objects, and the collision location relative to the Roche limit. We use either an ``angle-averaged'' criterion, that assumes randomly oriented collisions, and a ``total accretion'' criterion, in which collisions are assumed to occur in the radial direction along the widest axis of the Hill sphere. These are the same accretion criteria as those used in \citet{ida97}, and in some of the simulations in \citet{kokubo00}. Additional details can be found in Appendix \ref{appendix_tidal_accretion}.

While an improvement over the assumption of perfect mergers during every collision, our tidal accretion criteria are still idealized. They ignore the potential for fragmentation or substantial deformation when calculating whether a given collision results in accretion, and assume that an accreted pair merges into a new spherical body. \citet{kokubo00} considered three different accretion models: the two described above, and a \textquotedblleft rubble pile model\textquotedblright, in which individual N-body particles are never merged but allowed to form gravitationally bound aggregates of irregular shapes that can, e.g., be tidally disrupted when they pass within the Roche limit. \citet{kokubo00} find similar overall outcomes for all three treatments (e.g., their Figure 1); this is probably because the Moon's final position is affected more by its resonant interactions with the inner disk in their simulations than the exact position at which it begins to grow, so long as the latter is outside the Roche limit.  We use the angle-averaged criterion for direct comparison with Ida et al. (1997) in our Section 3 simulations, and the total accretion criterion in our hybrid simulations in Section 4.

\subsubsection{Tidal disruption of moonlets}
Close encounters between particles can lead to some of them being scattered toward the planet on high eccentricity orbits, where they may suffer tidal disruption and be effectively absorbed by the inner disk. We expect objects accreting in the outer disk to be molten or partially molten (e.g., Section \ref{section_physical_state}). An inviscid fluid object on a parabolic orbit will tidally disrupt in a single pass if its pericenter distance Q satisfies
\begin{equation}
 Q < 1.05\left(\frac{M_p}{\rho_0}\right)^{1/3},
\end{equation}
where $M_p$ is the mass of the disrupting body and $\rho_0$ is the density of the orbiting body \citep{sridhar92}. For the Earth-moon system, this yields $Q < 2R_\Earth$.

At each time step, we compute the distance of each object to the primary. If this distance is smaller than $2R_\Earth$, we remove the body from the N-body code and add its mass and angular momentum to that of the inner disk. The latter is done by finding the disk's new outer edge $r_d'$ so that $L_d-L_d'+L_c=0$, where $L_d$ and $L_d'$ are the disk's angular momentum before and after capture, and $L_c$ is the angular momentum of the captured body. The disk mass after capture is $M_d'=M_d+m_c$ where $m_c$ is the mass of the captured body. This yields
\begin{equation}
 \frac{4}{5}M_d\frac{r_d^{5/2}-R^{5/2}}{r_d^2-R^2}-\frac{4}{5}M_d'\frac{r_d'^{5/2}-R^{5/2}}{r_d'^2-R^2}
+m_c\sqrt{a_c\left(1-e_c^2\right)}=0,
\end{equation}
where $R=R_\Earth$ is the disk's inner edge (see Appendix \ref{appendix_disk_mass} for details), and $a_c$ and $e_c$ are the captured body's semi-major axis and eccentricity. We solve this numerically so that angular momentum is conserved to a $10^{-8}$ precision.

To prevent the inner disk's outer edge from expanding too far beyond the Roche limit in a single time step due to the tidal disruption of a large object, we implement a tidal stripping mechanism for large objects. If a body's mass is greater than $10^{-5} M_\Earth$, we remove 20\% of its mass at each time step once its distance to the primary is $r < 2R_\oplus$, so that large bodies are entirely disrupted over a few time steps. Since the time step in Symba is $1/20^{th}$ of the orbital period at 1 Earth radius, large bodies are then effectively disrupted over a timescale $< 30$ minutes, which is much shorter than their orbital period. On some runs, this still leads to the outer disk edge temporarily expanding to $\sim2.93R_\Earth$, but then new bodies are formed from fragmentation (see previous section), and the disk outer edge returns to close to the Roche limit in a few tens of orbits. Generally this happens late in simulations, when the disk mass is $\le 10^{-1} M_\leftmoon$, so we believe it does not greatly affect the outcome of our 
simulations.

We note that the orbits of bodies passing through the inner disk would also be affected by drag interaction with inner disk material. For example, an object that encounters a mass comparable to its own during a single passage through the inner disk would be captured by the inner disk.  We neglect this process here, since it depends sensitively on the disk properties (notably its scale height and radial surface density profile), which are treated in a simplified fashion by our model.

\section{Tests with pure N-body simulations}
We begin by performing pure N-body simulations, using the angle-average accretion criterion, for direct comparison with previous results of \citet{ida97}, using initial disk parameters given in Table 1 of that paper and summarized in our Table \ref{table_Nbody_dataset}.

For runs 1 to 14, the initial disk mass is $M_d = 2.44 M_{\leftmoon}$, and the index of the particle size distribution, $\left(N(m)dm\propto m^{-p} dm\right)$ is $p=1.5$. Different values are used for the disk's outer edge $a_{max}$, the surface density distribution exponent $q$ $\left(\sigma(a) \propto a^{-q}\right)$, and the number of particles $N$. For runs 15 to 19, the disk's outer edge and exponent of the surface density distribution $q$ are held constant, while $M_d$ and $p$ are varied. We run each simulation for $5000T_K$ where $T_K$ is the orbital period at one Earth radius. This timescale is equivalent to that of \citet{ida97}, who use a simulation time of $1000 T_K'$, with $T_K'$ being the orbital period at $a_R$. 

Particles are distributed randomly throughout the disk, with initial eccentricities and inclinations (in radians) of order $O\left(10^{-1}\right)$ as in \citet{ida97}. We use their values for the normal and tangential coefficients of restitution, with $\epsilon_n=0.01\text{ or }0.5$ and $\epsilon_t=1$. Damping only the normal component of the relative velocity of colliding particles can lead to situations in which two particles remain in close physical contact but do not actually merge by our accretion criteria, since the tangential component of their relative velocity remains unchanged.  When we detect such a situation, which can cause the simulation's timestep to become prohibitively small, we force the merging of the two particles. In practice, in a simulation with $\approx 2000$ initial particles, this occurs between 0 and 2 times.

\begin{deluxetable}{c c c c c c c c c c}
\tabletypesize{\footnotesize}
\tablewidth{0pt}
\tablecolumns{10}
\tablecaption{N-body simulations parameters. \label{table_Nbody_dataset}}
\tablehead{ & \colhead{$L_d/M_d$} & \colhead{$L_d$} & \colhead{$M_d$} & \colhead{$a_{max}$}\\
\colhead{Run} & \colhead{$\left(\sqrt{GM_\Earth a_R}\right)$} & \colhead{$\left(L_{EM}\right)$} & \colhead{$\left(M_\leftmoon\right)$} & \colhead{$\left(a_R\right)$} & \colhead{$p$} & \colhead{$q$} & \colhead{$N$} & \colhead{$\epsilon_n$} & \colhead{$\epsilon_t$}}
\startdata
1   & 0.670 & 0.295 & 2.44 & 0.95 & 1.5 & 5 & 1500 & 0.01 & 1\\
2   & 0.670 & 0.295 & 2.44 & 0.95 & 1.5 & 5 & 1500 & 0.5  & 1\\
3   & 0.690 & 0.304 & 2.44 & 0.95 & 1.5 & 4 & 1000 & 0.01 & 1\\
4   & 0.692 & 0.305 & 2.44 & 0.95 & 1.5 & 4 & 2000 & 0.01 & 1\\
5   & 0.722 & 0.318 & 2.44 & 1.25 & 1.5 & 3 & 1000 & 0.01 & 1\\
6   & 0.722 & 0.318 & 2.44 & 1.25 & 1.5 & 3 & 1000 & 0.5  & 1\\
7   & 0.767 & 0.338 & 2.44 & 1.25 & 1.5 & 3 & 1500 & 0.01 & 1\\
8   & 0.794 & 0.350 & 2.44 & 1.25 & 1.5 & 3 & 2700 & 0.01 & 1\\
9   & 0.813 & 0.358 & 2.44 & 1.50 & 1.5 & 2 & 1500 & 0.01 & 1\\
10  & 0.823 & 0.363 & 2.44 & 1.50 & 1.5 & 2 & 1000 & 0.5  & 1\\
11  & 0.834 & 0.367 & 2.44 & 1.80 & 1.5 & 2 & 1000 & 0.01 & 1\\
12  & 0.891 & 0.393 & 2.44 & 2.00 & 1.5 & 2 & 1000 & 0.01 & 1\\
13  & 0.958 & 0.422 & 2.44 & 2.00 & 1.5 & 1 & 1000 & 0.01 & 1\\
14  & 0.977 & 0.430 & 2.44 & 2.00 & 1.5 & 1 & 1000 & 0.01 & 1\\
15  & 0.738 & 0.325 & 2.44 & 1.25 & 0.5 & 3 & 1000 & 0.01 & 1\\
16  & 0.757 & 0.443 & 3.24 & 1.25 & 1.5 & 3 & 1000 & 0.5  & 1\\
17  & 0.767 & 0.449 & 3.24 & 1.25 & 1.5 & 3 & 1500 & 0.01 & 1\\
18  & 0.768 & 0.338 & 2.44 & 1.25 & 1   & 3 & 1000 & 0.01 & 1\\
19  & 0.778 & 0.228 & 1.62 & 1.25 & 1.5 & 3 & 1000 & 0.01 & 1\\
\enddata
\tablecomments{Input parameters for our pure N-body simulations, adapted from \citet{ida97}. $M_d$, $L_d$, and $a_{max}$ are the disk's initial mass, angular momentum, and outer edge. $L_d/M_d$ is the so-called specific angular momentum (in units of $\sqrt{GM_{\Earth}a_R}$). $p$ and $q$ are the exponents for the initial particle size distribution $\left(N(m)\propto m^{-p}\right)$ and surface density distribution $\left(\sigma(a) \propto a^{-q}\right)$, and $N$ is the initial number of particles in the disk. $\epsilon_n$ and $\epsilon_t$ are the normal and tangential coefficients of restitution (see text for discussion). Units of mass, distance and angular momentum are the present lunar mass $M_{\leftmoon}$, the Roche limit for silicates $a_R \approx 2.9 R_{\Earth}$, and the angular momentum of the Earth-Moon system $\left(L_{EM} = \unit{3.5\times10^{41}}{\gram\usk\centi\squaremetre\usk\reciprocal\second}\right)$.}
\end{deluxetable}

\begin{deluxetable}{c c c c c c c c c c c c c}
\tabletypesize{\footnotesize}
\tablewidth{0pt}
\tablecolumns{12}
\tablecaption{N-body simulations results. \label{table_Nbody_results}}
\tablehead{ & \colhead{$a$} & & \colhead{$M$} & \colhead{$a_2$} & \colhead{$M_2$} & \colhead{$M'$} & \colhead{$M_{pl}$} & \colhead{$M_{\infty}$} & \colhead{$L$} & \colhead{$L'$} & \colhead{$L_{pl}$} & \colhead{$L_{\infty}$}\\
\colhead{Run} & \colhead{$\left(a_R\right)$} & \colhead{$e$} & \colhead{$\left(M_\leftmoon\right)$} & \colhead{$\left(a_R\right)$} & \colhead{$\left(M_\leftmoon\right)$} & \colhead{$\left(M_\leftmoon\right)$} & \colhead{$\left(M_\leftmoon\right)$} & \colhead{$\left(M_\leftmoon\right)$} & \colhead{$\left(L_{EM}\right)$} & \colhead{$\left(L_{EM}\right)$} & \colhead{$\left(L_{EM}\right)$} & \colhead{$\left(L_{EM}\right)$}}
\startdata
1   & 1.54 & 0.02 & 0.238 & 0.60 & 0.011 & 0.239 & 2.160 & 0.000 & 0.053 & 0.054 & 0.236 & 0.000 \\
2   & 0.91 & 0.06 & 0.120 & 1.99 & 0.047 & 0.286 & 2.152 & 0.002 & 0.021 & 0.062 & 0.233 & 0.001 \\
3   & 1.43 & 0.04 & 0.367 & 0.57 & 0.019 & 0.369 & 2.036 & 0.002 & 0.079 & 0.079 & 0.222 & 0.001 \\
4   & 1.21 & 0.07 & 0.162 & 2.06 & 0.149 & 0.322 & 2.045 & 0.000 & 0.032 & 0.074 & 0.221 & 0.000 \\
5   & 1.44 & 0.03 & 0.424 & 6.54 & 0.017 & 0.442 & 1.956 & 0.005 & 0.092 & 0.097 & 0.216 & 0.002 \\
6   & 0.81 & 0.16 & 0.139 & 1.45 & 0.081 & 0.416 & 2.012 & 0.012 & 0.022 & 0.095 & 0.222 & 0.003 \\
7   & 1.23 & 0.06 & 0.605 & 0.61 & 0.072 & 0.612 & 1.693 & 0.052 & 0.121 & 0.123 & 0.186 & 0.015 \\
8   & 1.32 & 0.06 & 0.734 & 0.63 & 0.058 & 0.737 & 1.613 & 0.031 & 0.152 & 0.153 & 0.176 & 0.010 \\
9   & 1.50 & 0.14 & 0.545 & 0.58 & 0.179 & 0.611 & 1.588 & 0.061 & 0.119 & 0.140 & 0.176 & 0.020 \\
10  & 1.19 & 0.07 & 0.654 & 8.93 & 0.015 & 0.774 & 1.591 & 0.065 & 0.128 & 0.165 & 0.179 & 0.021 \\
11  & 1.38 & 0.04 & 0.827 & 3.84 & 0.002 & 0.830 & 1.558 & 0.051 & 0.175 & 0.176 & 0.176 & 0.017 \\
12  & 0.78 & 0.28 & 0.517 & 2.25 & 0.505 & 1.045 & 1.277 & 0.119 & 0.079 & 0.222 & 0.144 & 0.035 \\
13  & 1.57 & 0.07 & 0.828 & 0.96 & 0.381 & 0.845 & 1.096 & 0.118 & 0.187 & 0.192 & 0.125 & 0.040 \\
14  & 1.52 & 0.04 & 1.315 & 0.66 & 0.077 & 1.315 & 0.982 & 0.066 & 0.292 & 0.292 & 0.111 & 0.023 \\
15  & 1.33 & 0.04 & 0.537 & 0.66 & 0.013 & 0.551 & 1.850 & 0.013 & 0.111 & 0.116 & 0.202 & 0.004 \\
16  & 1.25 & 0.02 & 0.525 & 0.67 & 0.016 & 0.663 & 2.487 & 0.063 & 0.106 & 0.146 & 0.275 & 0.020 \\
17  & 1.40 & 0.05 & 0.768 & 0.56 & 0.093 & 0.768 & 2.321 & 0.056 & 0.164 & 0.164 & 0.256 & 0.018 \\
18  & 1.65 & 0.05 & 0.496 & 0.84 & 0.029 & 0.504 & 1.885 & 0.010 & 0.115 & 0.117 & 0.213 & 0.004 \\
19  & 1.56 & 0.06 & 0.411 & 0.89 & 0.018 & 0.413 & 1.165 & 0.000 & 0.092 & 0.093 & 0.129 & 0.000 \\
\enddata
\tablecomments{$a$, $e$ $M$, and $L$ are the semi-major axis, eccentricity, mass, and angular momentum of the largest moon at the end of the simulation ($t=5000 T_K$). $a_2$ and $M_2$ are the semi-major axis and mass of the second largest body. $M'$ and $L'$ are the mass and angular momentum of the largest body plus all bodies outside of its orbit. $M_{pl}$ and $L_{pl}$ are the mass and angular momentum of particles that were scattered onto the planet. $M_{\infty}$ and $L_{\infty}$ are the mass and angular momentum of ejected particles. Units of mass, distance and angular momentum are the present lunar mass $M_{\leftmoon}$, the Roche limit for silicates $a_R \approx 2.9 R_{\Earth}$, and the angular momentum of the Earth-Moon system $\left(L_{EM} = \unit{3.5\times10^{41}}{\gram\usk\centi\squaremetre\usk\reciprocal\second}\right)$.}
\end{deluxetable}

The most common outcome of the \citet{ida97} simulations was a single large Moon with an average semi-major axis of $\langle a\rangle \approx 1.3a_R$.  Because inner disk particles spread rapidly and are strongly scattered by the forming Moon, all or nearly all of the disk material interior to the Roche limit in their simulations was removed in less than a year (through collisions with the Earth or Moon, or by escape from the system). A primary finding of \citet{ida97}  \citep[see also][]{kokubo00} was that the fraction of the initial disk's mass incorporated into the final Moon is a function of the initial specific angular momentum of the disk ($L_d/M_d$) and the fraction of the disk that escapes during the Moon's accretion ($M_{\infty}/M_d$).  They used conservation of mass and angular momentum to analytically estimate the mass of the largest Moon as:
\begin{equation}
{\frac{M}{M_d}}\approx \frac{1.9 L_d}{M_d\sqrt{GM_{\Earth}a_R}} - 1.15 -1.9 \frac{M_\infty}{M_d},
\label{equ_MoonMassRelation}
\end{equation}
where they assumed that the Moon forms at $a =1.3a_R$ (see also Section \ref{section_analytical_estimate}).

\begin{figure}[!h]
\begin{center}
\includegraphics[width=8cm]{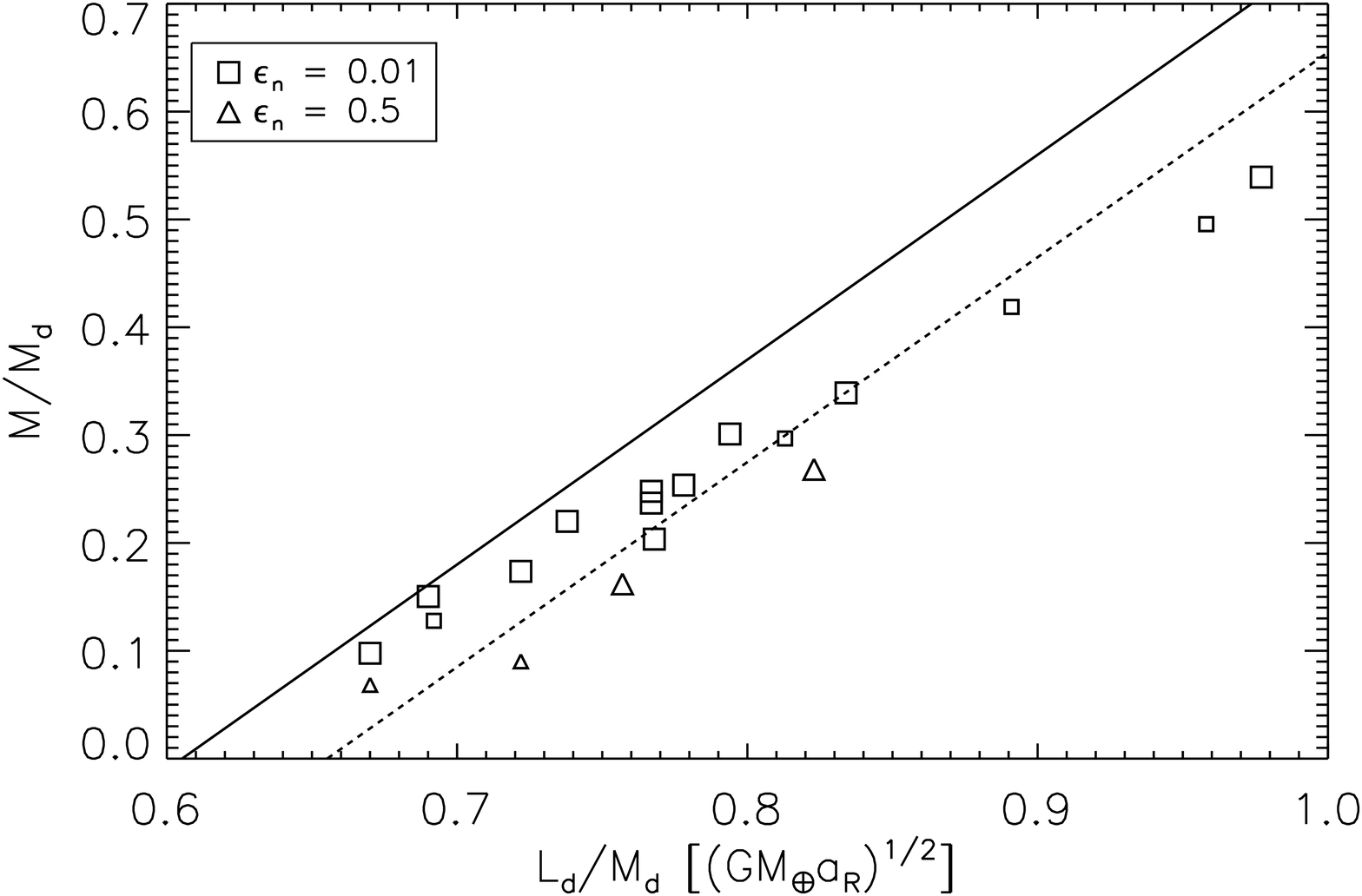}
\end{center}
\caption{Ratio of the mass of the largest body at $t=5000 T_K$ to the initial disk mass, as a function of the disk's initial specific angular momentum, for pure N-body simulations comparable to those in Ida et al. (1997). Squares correspond to Runs with $\epsilon_n=0.01$ and triangles to those with $\epsilon_n=0.5$. Small symbols are cases where the mass of the second largest body is at least $30\%$ that of the largest one, in which cases we plotted the mass of the combined bodies. The solid and dashed lines corresponds to equation \ref{equ_MoonMassRelation} with $M_\infty = 0$ and $M_\infty = 0.05 M_d$, respectively.}
\label{fig_Nbody_results}
\end{figure}

Our pure N-body simulations generally reproduce the findings of \citet{ida97}. In Figure \ref{fig_Nbody_results} we plot the ratio of the mass of the largest orbiting body at $t=5000 T_K$ to the initial disk mass for our simulations, as a function of the disk's initial specific angular momentum. The results we obtain are similar to those shown in Figure 5 of \citet{ida97}. For the most extended disks (Runs 13 and 14) we find somewhat larger objects than in \citet{ida97}, although our results for these cases are similar to those obtained by \citet{kokubo00} for comparable initial $L_d/M_d$ values. Analytical estimates from equation \ref{equ_MoonMassRelation} with $M_\infty = 0$ (solid line) and $M_\infty = 0.05 M_d$ (dashed line) are also plotted on Figure \ref{fig_Nbody_results}.  As in \citet{ida97} and \citet{kokubo00}, we find that $M_\infty$ increases for initially more radially extended disks (i.e. for disks with larger $L_d/M_d$).  The ratio of the Moon's escape velocity to the local escape velocity from the Earth is $(2GM_{\leftmoon}/R_{\leftmoon})^{1/2}
/(2GM_{\oplus}/a)^{1/2} \approx 0.4(a/a_R)^{1/2}$, so that lunar-sized objects are increasingly effective at gravitationally scattering particles into escaping orbits as their orbital radii increase. 

The average semi-major axis, eccentricity, and mass of the final largest moons in our simulations are $\langle a\rangle = 1.32 a_R$, $\langle e\rangle = 0.07$, and $\langle M\rangle = 0.54M_{\leftmoon}$, in good agreement with $\langle a\rangle = 1.27 a_R$, $\langle e\rangle < 0.1$, and $\langle M\rangle = 0.48 M_{\leftmoon}$ from \citet{ida97}.

\section{Simulations with a Roche-interior fluid disk}
We here describe our hybrid simulations that model the Roche-interior disk as a fluid and material exterior to the Roche limit with individual particles. For these simulations we adopt the total accretion criterion, for which accretion is possible between like-sized objects for $a \geq a_R$ (see Appendix \ref{appendix_tidal_accretion}).

\subsection{Simulation parameters}
\begin{figure}[!h]
\begin{center}
\includegraphics[width=8cm]{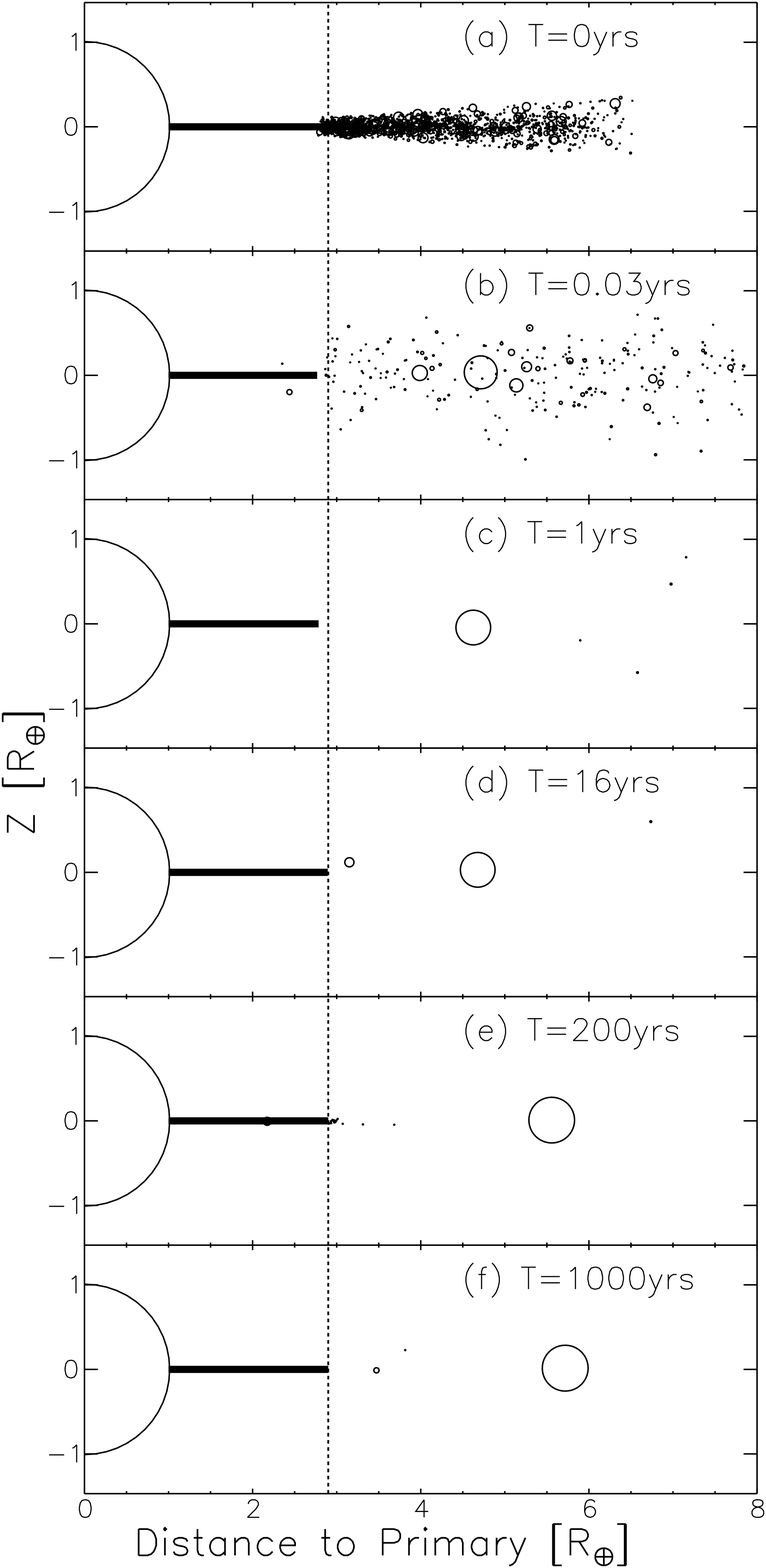}
\end{center}
\caption{Snapshots of the protolunar disk, projected on the $R-z$ plane, at $t=0$, 0.03, 1, 30, 200 and $\unit{1000}{years}$, for Run 34 using the hybrid model with a fluid inner disk. The size of circles is proportional to the physical size of the corresponding particle. The horizontal thick line is the Roche-interior disk. The vertical dashed line is the Roche limit at $2.9 R_{\Earth}$.}
\label{fig_simulation_snapshots}
\end{figure}

Recent impact simulations suggest that the protolunar disk had a mass of $1.5 - 2.1 M_{\leftmoon}$ and a specific angular momentum of $0.8 - 1.1$, in units of $\sqrt{GM_{\Earth}a_R}$ \citep{canup12}. Since we adopt a uniform surface density for the Roche-interior disk, the minimum specific angular momentum we can achieve, corresponding to a case with only an inner fluid disk extending from 1 to $2.9 R_{\Earth}$ is $L_d/M_d \approx 0.845\sqrt{GM_{\Earth}a_R}$ from equation (\ref{equ_InnerDiskAngularMomentum}). Simulation parameters are shown in Table \ref{table_Nbodyandfluid_dataset}. We consider cases with initial total disk masses $M_d=2$, 2.4, 2.5 and $3 M_{\leftmoon}$; inner disk masses, $M_{in}$, that contain between 50\% and 100\% of the total disk mass; outer disk edges (global) $a_{max}=2.9$, 4, 6, 7 and $8 a_R$; and exponents for the surface density distribution in the outer disk of $q=1, 3\text{ and }5$. We fix the particle size distribution exponent at $p=1.5$, the normal and tangential coefficients of restitution $\epsilon_n=0.01$ and $\epsilon_t=1$, and the number of particles $N=1500$, as those proved to be of little influence in pure N-body simulations. An example of the initial setup is plotted in Figure \ref{fig_simulation_snapshots}a, for Run 34.

\begin{deluxetable}{c c c c c c c c}
\tabletypesize{\footnotesize}
\tablewidth{0pt}
\tablecolumns{8}
\tablecaption{Hybrid simulations parameters. \label{table_Nbodyandfluid_dataset}}
\tablehead{ & \colhead{$L_d/M_d$} & \colhead{$L_d$} & \colhead{$M_d$} & \colhead{$M_{in}$} & \colhead{$M_{out}$} & & \colhead{$a_{max}$}\\
\colhead{Run} & \colhead{$\left(\sqrt{GM_\Earth a_R}\right)$} & \colhead{$\left(L_{EM}\right)$} & \colhead{$\left(M_\leftmoon\right)$} & \colhead{$\left(M_\leftmoon\right)$} & \colhead{$\left(M_\leftmoon\right)$} & \colhead{$q$} & \colhead{$\left(R_\Earth\right)$}}
\startdata
1  & 0.843 & 0.304 & 2.00 & 2.00 & 0.00 & N/A & 2.9\\
2  & 0.843 & 0.365 & 2.50 & 2.50 & 0.00 & N/A & 2.9\\
\hline
3  & 0.955 & 0.345 & 2.00 & 1.00 & 1.00 & 5   & 4\\
4  & 0.960 & 0.347 & 2.00 & 1.00 & 1.00 & 3   & 4\\
5  & 0.965 & 0.348 & 2.00 & 1.00 & 1.00 & 1   & 4\\
6  & 0.955 & 0.414 & 2.40 & 1.20 & 1.20 & 5   & 4\\
7  & 0.960 & 0.416 & 2.40 & 1.20 & 1.20 & 3   & 4\\
8  & 0.965 & 0.418 & 2.40 & 1.20 & 1.20 & 1   & 4\\
9  & 0.899 & 0.325 & 2.00 & 1.50 & 0.50 & 5   & 4\\
10 & 0.901 & 0.326 & 2.00 & 1.50 & 0.50 & 3   & 4\\
11 & 0.904 & 0.326 & 2.00 & 1.50 & 0.50 & 1   & 4\\
12 & 0.899 & 0.390 & 2.40 & 1.80 & 0.60 & 5   & 4\\
13 & 0.901 & 0.391 & 2.40 & 1.80 & 0.60 & 3   & 4\\
14 & 0.904 & 0.392 & 2.40 & 1.80 & 0.60 & 1   & 4\\
15 & 0.888 & 0.401 & 2.50 & 2.00 & 0.50 & 5   & 4\\
16 & 0.890 & 0.402 & 2.50 & 2.00 & 0.50 & 3   & 4\\
17 & 0.892 & 0.403 & 2.50 & 2.00 & 0.50 & 1   & 4\\
18 & 0.880 & 0.477 & 3.00 & 2.50 & 0.50 & 5   & 4\\
19 & 0.882 & 0.478 & 3.00 & 2.50 & 0.50 & 3   & 4\\
20 & 0.884 & 0.479 & 3.00 & 2.50 & 0.50 & 1   & 4\\
\hline
21 & 0.986 & 0.356 & 2.00 & 1.00 & 1.00 & 5   & 6\\
22 & 1.009 & 0.365 & 2.00 & 1.00 & 1.00 & 3   & 6\\
23 & 1.036 & 0.374 & 2.00 & 1.00 & 1.00 & 1   & 6\\
24 & 0.986 & 0.427 & 2.40 & 1.20 & 1.20 & 5   & 6\\
25 & 1.009 & 0.437 & 2.40 & 1.20 & 1.20 & 3   & 6\\
26 & 1.036 & 0.449 & 2.40 & 1.20 & 1.20 & 1   & 6\\
27 & 0.914 & 0.330 & 2.00 & 1.50 & 0.50 & 5   & 6\\
28 & 0.926 & 0.335 & 2.00 & 1.50 & 0.50 & 3   & 6\\
29 & 0.940 & 0.339 & 2.00 & 1.50 & 0.50 & 1   & 6\\
30 & 0.914 & 0.396 & 2.40 & 1.80 & 0.60 & 5   & 6\\
31 & 0.926 & 0.401 & 2.40 & 1.80 & 0.60 & 3   & 6\\
32 & 0.940 & 0.407 & 2.40 & 1.80 & 0.60 & 1   & 6\\
33 & 0.900 & 0.406 & 2.50 & 2.00 & 0.50 & 5   & 6\\
34 & 0.909 & 0.411 & 2.50 & 2.00 & 0.50 & 3   & 6\\
35 & 0.920 & 0.416 & 2.50 & 2.00 & 0.50 & 1   & 6\\
36 & 0.890 & 0.482 & 3.00 & 2.50 & 0.50 & 5   & 6\\
37 & 0.898 & 0.487 & 3.00 & 2.50 & 0.50 & 3   & 6\\
38 & 0.907 & 0.492 & 3.00 & 2.50 & 0.50 & 1   & 6\\
\hline
39 & 1.068 & 0.386 & 2.00 & 1.00 & 1.00 & 1   & 7\\
40 & 1.068 & 0.463 & 2.00 & 1.20 & 1.20 & 1   & 7\\
%\hline
%\tablebreak
41 & 0.998 & 0.361 & 2.00 & 1.00 & 1.00 & 5   & 8\\
42 & 1.043 & 0.377 & 2.00 & 1.00 & 1.00 & 3   & 8\\
43 & 1.099 & 0.397 & 2.00 & 1.00 & 1.00 & 1   & 8\\
44 & 0.998 & 0.433 & 2.40 & 1.20 & 1.20 & 5   & 8\\
45 & 1.043 & 0.452 & 2.40 & 1.20 & 1.20 & 3   & 8\\
46 & 1.098 & 0.476 & 2.40 & 1.20 & 1.20 & 1   & 8\\
\hline
\enddata
\tablecomments{Simulation parameters with a Roche-interior fluid disk and Roche-exterior individual particles. $M_d$, $L_d$, and $a_{max}$ are the disk's total initial mass, angular momentum, and outer edge. $M_{in}$ and $M_{out}$ are the masses of the fluid disk, and of the solid bodies, respectively. $L_d/M_d$ is the disk's total specific angular momentum (in units of $\sqrt{GM_{\Earth}a_R}$). $q$ is the exponent for the initial surface density distribution $\left(\sigma(a) \propto a^{-q}\right)$ of the Roche-exterior disk. Units of mass, distance and angular momentum are the present lunar mass $M_{\leftmoon}$, Earth radius $R_\Earth$, and angular momentum of the Earth-Moon system $\left(L_{EM} = \unit{3.5\times10^{41}}{\gram\usk\centi\squaremetre\usk\reciprocal\second}\right)$. The normal and tangential coefficients of restitution $\epsilon_n$ and $\epsilon_t$ are set to 0.01 and 1, respectively. The particle-size distribution index $p$ is set to 1.5, and the number of 
orbiting particles $N$ is set to 1500. Runs 1 and 2 start with only a Roche-interior fluid disk.}
\end{deluxetable}

\subsection{Accretion dynamics}\label{section_accretion_dynamics}
\subsubsection{A three-stage accretion}\label{section_three_stage_accretion}
\begin{figure}[!h]
\begin{center}
\includegraphics[width=8cm]{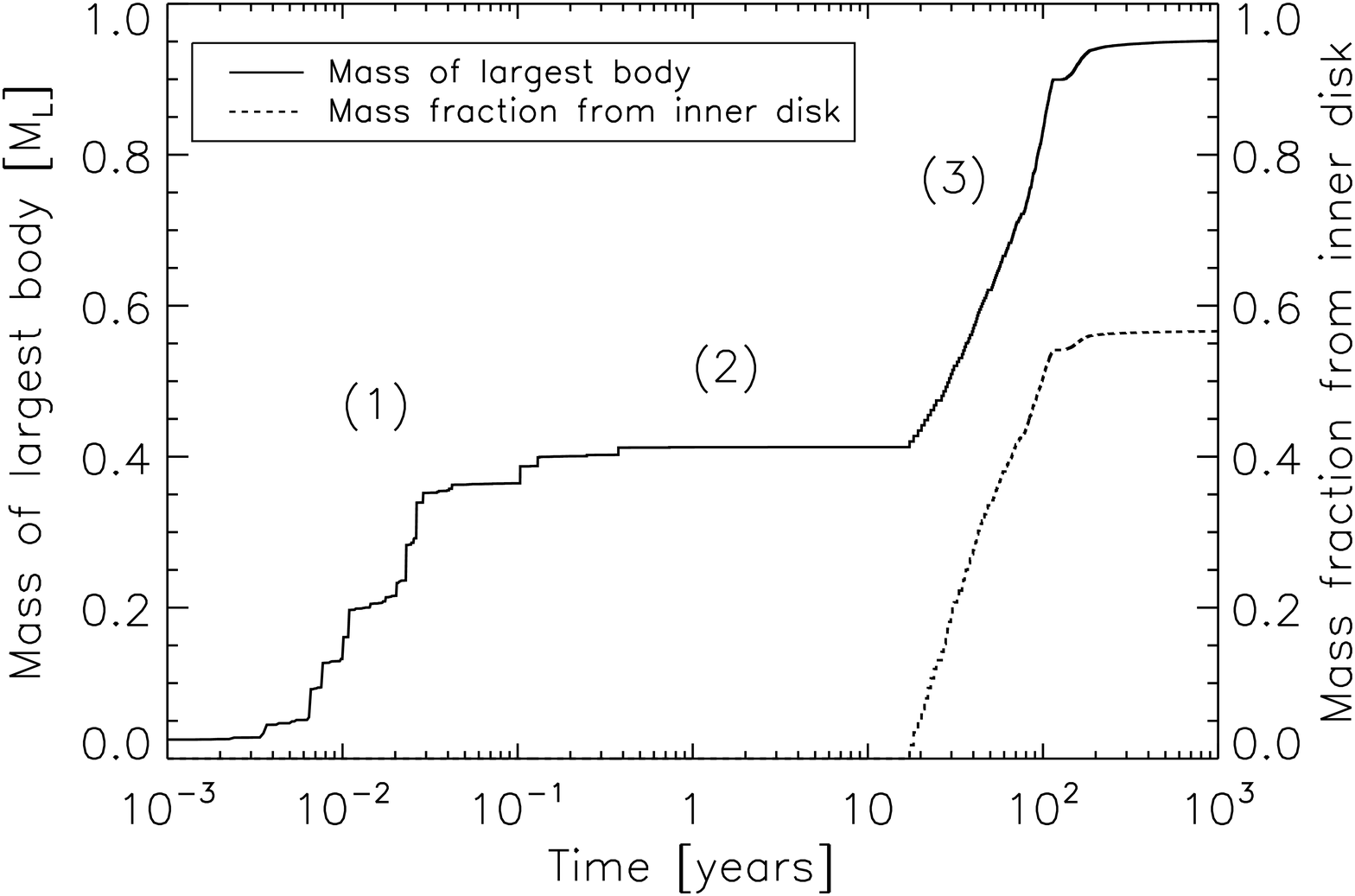}
\end{center}
\caption{Mass of the largest body in Run 34 (solid line), and fraction of its mass composed of material derived from the Roche-interior disk (dashed line). First Roche-exterior bodies collide and accrete until only a few massive bodies remain (1). These bodies confine the inner disk due to resonant interactions, and in turn they recede away (2). Eventually, the inner disk viscously spreads back out to the Roche limit, and new moonlets are spawned that collide with the Moon and complete its growth (3).}
\label{fig_mass_and_fraction_largest}
\end{figure}

\begin{figure}[!h]
\begin{center}
\includegraphics[width=8cm]{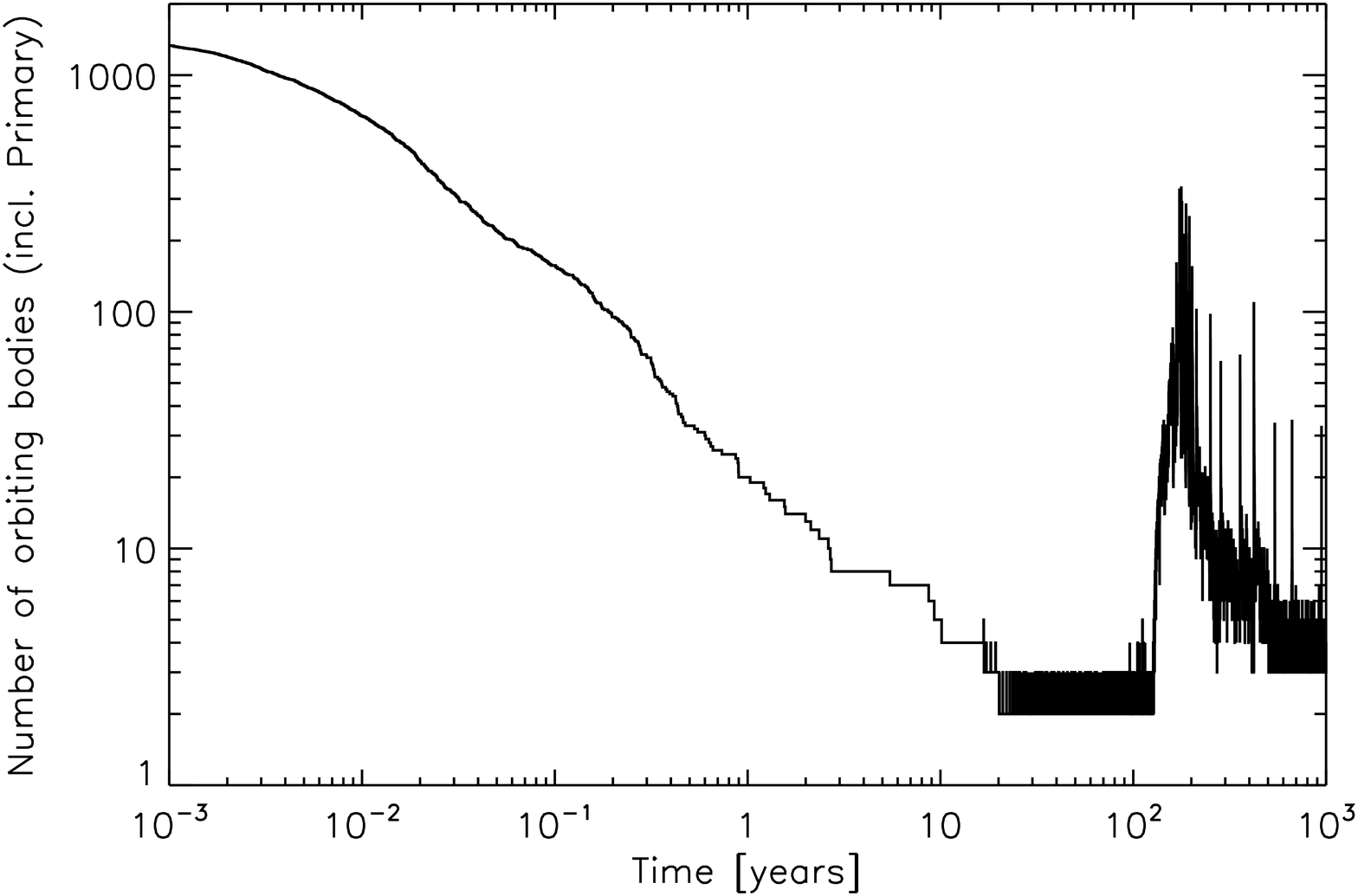}
\end{center}
\caption{Number of orbiting bodies in Run 34. Most bodies initially present in the outer disk merge or get scattered in $\approx \unit{1}{year}$. After $\approx \unit{20}{years}$, the Roche-interior disk has respread back out to the Roche limit and starts producing new moonlets as material spreads beyond $a_R$. After $\approx \unit{200}{years}$, the disk produces fragments that are very small, get captured in the 2:1 mean motion resonance with the Moon, and are mostly scattered onto the planet (see details in Section \ref{section_migration}).}
\label{fig_n_orbiting_bodies}
\end{figure}

Figure \ref{fig_mass_and_fraction_largest} shows the evolution of the mass of the largest body in Run 34 (solid line), as well as the fraction of its mass that consists of material accreted from the Roche-interior disk (dashed line). Figure \ref{fig_n_orbiting_bodies} shows the evolution of the number of orbiting bodies. 

The accretion of the Moon occurs in three consecutive phases: (1) Roche-exterior bodies rapidly collide, accrete and scatter one another until only a few massive bodies remain after $\sim \unit{1}{year}$ (Figure \ref{fig_simulation_snapshots}b and c, Figure \ref{fig_mass_and_fraction_largest}). (2) The inner disk is confined due to resonant interactions with outer bodies, which in turn recede away as the inner disk slowly viscously spreads outward. During that time, the growth of the Moon is stalled, but the inner disk loses mass on the planet. (3) After $\sim \unit{20}{years}$, the inner disk spreads back out to the Roche limit, and new moonlets are spawned (Figure \ref{fig_simulation_snapshots}d). These new objects either collide with the Moon to continue its accretion, or get ejected or scattered close to the planet where they are absorbed by the inner disk.  When the Moon accretes spawned moonlets, its semi-major axis tends to decrease slightly, since the specific angular momentum of the spawned moonlets is typically smaller than that of the Moon.  However interactions between the Moon and moonlets that are scattered into the inner disk cause the Moon's semi-major axis to increase, as the Moon generally gains angular momentum from the inner scattered bodies (Figure \ref{fig_simulation_snapshots}e and f, see also next section).  The latter effect dominates the end of the system's evolution.

Contrary to accretion timescales of less than a year found with pure N-body simulations, here the initial confinement of the inner disk by outer bodies and the slow spreading of the Roche-interior disk back out to the Roche limit delay the final accretion of the Moon by several hundreds of years. We can estimate the minimum mass of an outer object capable of strongly confining the inner disk by setting a moon's resonant torque on the disk equal to the disk's viscous torque, assuming that a single object confines the inner disk via its 2:1 inner Lindblad resonance. The disk's viscous torque at its outer edge reads
\begin{equation}
 \Gamma_\nu=3\pi\nu\sigma r_d^2\Omega.
\end{equation}
Assuming that the inner disk contains $\sim 1 M_\leftmoon$, it will evolve with the radiation-limited viscosity $\nu=\nu_{TS}$ (see Section \ref{section_model_used}). 
The confining satellite's torque reads
\begin{equation}
 \Gamma_m=\frac{\pi^2M_s^2G\sigma a_sc_m}{3M_\Earth}.
\end{equation}
Assuming that the confining body's 2:1 resonance lies at the inner disk's outer edge, we have $r_d=\left(1-1/m\right)^{2/3}a_s \approx 0.63a_s$, and setting $\Gamma_m = \Gamma_{\nu}$ requires
\begin{equation}
\left(\frac{M_s}{M_\oplus}\right) \approx \left [ \left(\frac{m}{m-1}\right )^{1/3} \frac {1}{m^2} \left (\frac {\nu}{r_d^2\Omega}\right )\right ]^{1/2}.
\end{equation}

Strongly confining a $1M_\leftmoon$ disk with $r_d=a_R$ and $\nu = \nu_{TS}$ requires an outer moon with a mass $M_s \geq 0.07 M_\leftmoon$. Thus initially, relatively small moonlets can confine the disk because the radiation-limited viscosity is not very strong. Because this viscosity is inversely proportional to the disk's surface density, it becomes increasingly difficult to confine the disk as it becomes less massive over time so long as the disk is radiation-limited.  However once the disk mass drops to $\leq 0.2M_{\leftmoon}$, the viscosity changes to an instability-induced viscosity with $\nu = \nu_{WC} \propto {\sigma}^2$, and the disk then becomes progressively easier to confine as it dissipates.

\subsubsection{Moonlet-driven orbital migration}
\begin{figure}[!h]
\begin{center}
\includegraphics[width=8cm]{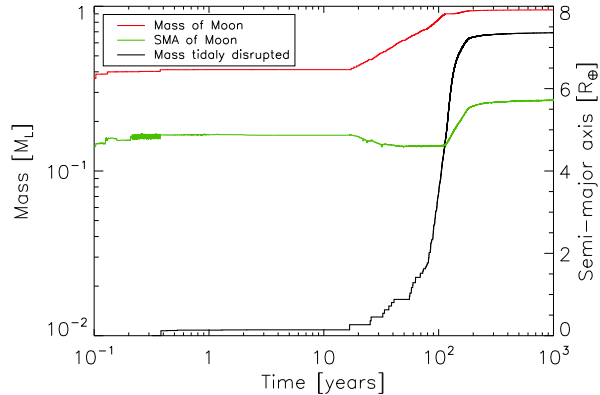}
\end{center}
\caption{Mass and semi-major axis (SMA) of the largest body, and cumulative mass of objects that were tidally disrupted and absorbed into the inner disk after being scattered close to the planet.}
\label{fig_sma_largest_body}
\end{figure}

{\label{section_migration}}
Figure \ref{fig_sma_largest_body} shows the evolution of the mass and semi-major axis of the largest body, and the cumulative mass of particles tidally disrupted and absorbed by the inner disk as they are scattered close to the planet, for Run 34. While the average semi-major axis of the Moon is $\sim 1.3 a_R$ in pure N-body simulations, with a Roche-interior fluid disk this value is increased to $\sim 2.15 a_R$ (see also Table \ref{table_Nbodyandfluid_results}). 

The Moon formed here in phase (1) by accretion of the initial outer bodies lies at $\sim 4.8 R_{\Earth}$. During that phase the inner disk's outer edge has been confined within $\sim 2.8 R_\Earth$, so that the Moon does not have any resonant interactions with the disk. The latter then slowly viscously spreads outward, until it reaches the Roche limit at $t\approx \unit{20}{years}$, at which point new moonlets are spawned.

Initially, the Moon efficiently accretes these moonlets, causing its mass to increase and its semi-major axis to decrease slightly (Figure \ref{fig_sma_largest_body}, red and green lines). This is due to a change in the Moon's angular momentum. Before accreting an object, the latter reads
\begin{equation}
L=M \sqrt{a G M_\earth \left(1-e^2\right)},
\end{equation}
where $M$, $a$ and $e$ are the moon's mass, semi-major axis and eccentricity. After accreting a fragment of mass $m_f$, the Moon's angular momentum reads
\begin{equation}
L'=\left( M+m_f\right) \sqrt{a' G M_\earth \left(1-e'^2\right)},
\end{equation}
where $a'$ and $e'$ are the post-accretion semi-major axis and eccentricity of the moon. Finally, the fragment's angular momentum is 
\begin{equation}
L_f=m_f \sqrt{a_f G M_\earth \left(1-e_f^2\right)}.
\end{equation}
Conservation of angular momentum gives $L+L_f = L'$. Since the Moon's eccentricity is generally of order $10^{-2} - 10^{-3}$, we can set $e^2 \approx e'^2 \approx 0$, which gives
\begin{equation}
a' \approx \left( \frac{M\sqrt{a} + m_f \sqrt{a_f\left(1-e_f^2\right)}}{M+m_f}\right)^2.
\end{equation}
Since $1-e_f^2 < 1$ and $a_f < a$, we get $a' < a$. The Moon's inward migration is however stopped at $\sim 4.6 R_\earth$, since when it goes inside that distance its 2:1 resonance falls into the disk, reactivating its disk torques and resulting in an outward migration of the Moon.

In order for a new moonlet to collide with the Moon, their orbits must cross. This can be done by an increase in the moonlet's semi-major axis and/or its eccentricity. However, if the latter occurs then the object can have a pericenter close enough to the planet that it would be tidally disrupted before encountering the Moon. A rapid increase of the moonlet's semi-major axis before its eccentricity gets too high is a more favorable scenario for a moonlet to succesfully collide with the Moon.

\begin{figure}[!h]
\begin{center}
\includegraphics[width=8cm]{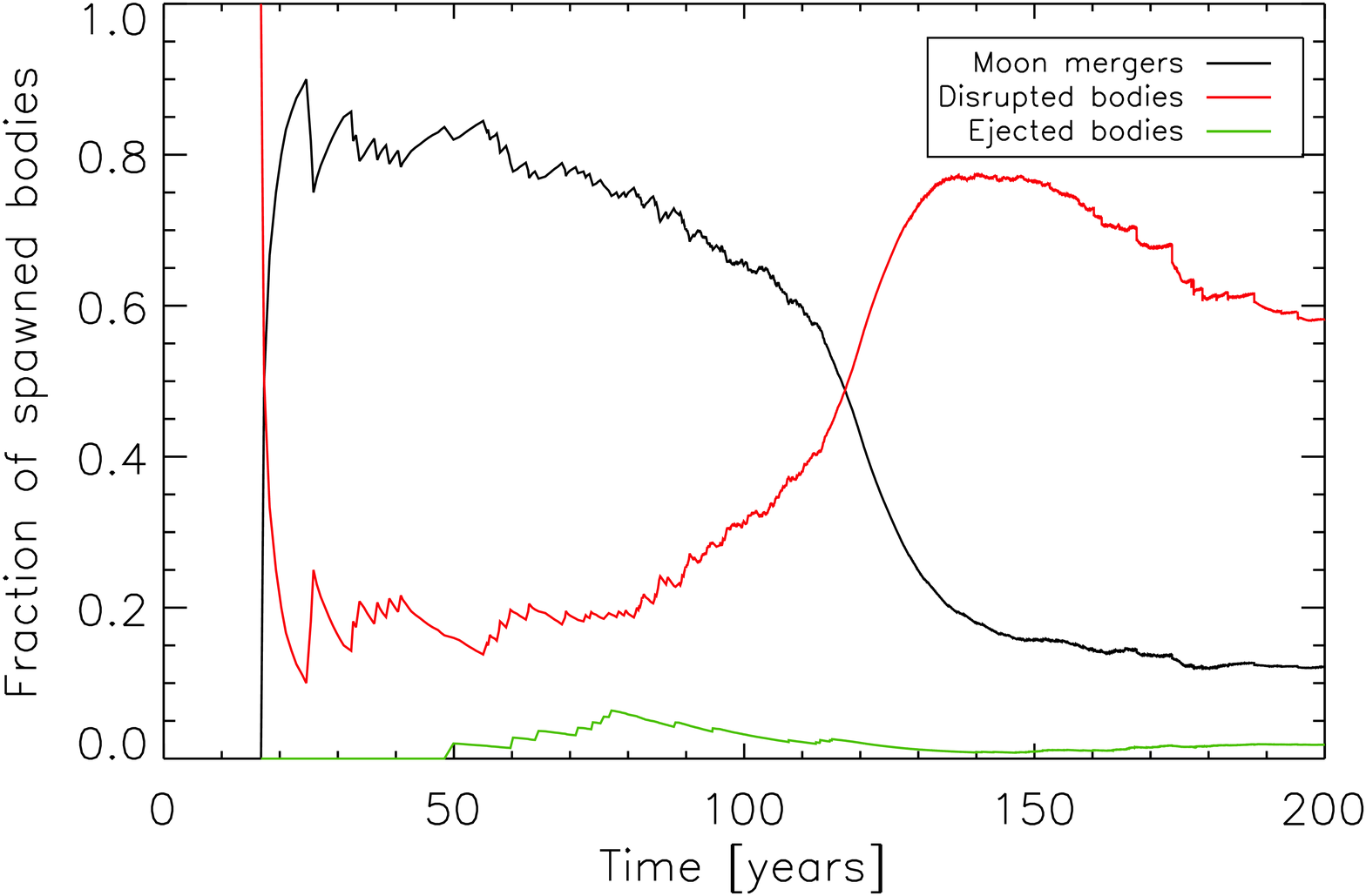}
\end{center}
\caption{Fraction of bodies spawned at the Roche limit that merge with the Moon (black), get tidally disrupted (red), or get ejected from the system (green). Initially most bodies merge with the Moon, but as the torque from the inner disk decreases due to its decreasing mass, progressively more bodies get scattered toward the planet and are tidally disrupted (see text for details). After $\approx \unit{150}{years}$ the total of the 3 curves is not equal to 1 because many bodies are trapped in the 2:1 Moon's resonance, at which point their ``fate'' has not yet been decided.}
\label{fig_spawned_particles_fate}
\end{figure}

The inner disk continuously loses mass on the planet and through the Roche limit, so that the torque it applies on newly formed objects decreases over time (see Eq. \ref{equ_disktorque_on_sat}). On the other hand, the Moon gets more massive over time, making it an even more efficient scatterer. Those two effects result in a progressively slower expansion of the semi-major axis of new moonlets, while their eccentricity gets excited even more rapidly by the growing Moon. As a result, it becomes increasingly difficult for new objects to collide with the Moon before getting tidally disrupted. Figure \ref{fig_spawned_particles_fate} shows the fraction of newly spawned objects that get accreted onto the Moon, tidally disrupted, or ejected from the system. It shows that initially, most of the new moonlets collide with the Moon. But as time goes by, a larger fraction of objects get scattered toward the planet and are tidally disrupted, until this outcome becomes predominant at $\sim \unit{120}{years}$.

A single interaction at conjunction between the Moon and an inner moonlet that is on an approximately circular orbit leads to a positive torque on the Moon's orbit.  Once the inner moonlet's orbit becomes eccentric, subsequent interactions between it and the Moon can lead to a positive or negative torque on the Moon's orbit. But if inner moonlets are removed by tidal disruption soon after their initial encounters with the Moon, the net torque on the Moon is on average positive, which drives an increase in its semi-major axis (Figure \ref{fig_sma_largest_body}, green line). When scattering events become predominant, the Moon starts migrating outward, at which point it becomes much more difficult for objects spawned at the Roche limit to merge directly with the Moon, and thus its growth levels off (Figure \ref{fig_sma_largest_body}, plateau on the red curve at $\sim \unit{120}{years}$).

When the Moon's 2:1 resonance is initially located just outside the disk's outer edge at $\approx 3R_\oplus$, spawned moonlets are not captured into the resonance. The latter requires that the change in a moonlet's semi-major axis due to an external torque in one libration period of the resonance be much less than the libration width of the resonance \citep[e.g.][]{dermott88}. This adiabatic condition is violated when the 2:1 is very near the disk edge, because the rate of increase in a moonlet's semi-major axis due to disk torques is typically too rapid as it crosses the resonance for inner disk masses $\geq 0.1M_\leftmoon$.  

However as the Moon's orbit expands outward due to scattering as described above, its 2:1 resonance moves away from the disk edge. The disk torque on a moonlet as it crosses the resonance is then weaker due to a greater separation between the moonlet and the disk's edge, and capture into the 2:1 resonance can occur as the disk is dissipating. For example, in Run 34, newly spawned moonlets begin to be trapped in the Moon's 2:1 at about $t = \unit{120}{years}$, when the 2:1 has moved outward to about $\unit{3.1}{R_\Earth}$, and the inner disk mass has decreased to $\approx 0.5 M_\leftmoon$. An inner moonlet trapped in the 2:1 resonance continues to receive a positive torque from the inner disk, but because it is in resonance with the Moon, the torque causes both the inner body's and the Moon's semi-major axes to expand in lock-step. In this way the Moon's orbit is driven outward due to indirect resonant interactions with the disk, with the inner moonlets acting as an angular momentum relay between the disk and the Moon.  Most moonlets captured into the Moon's 2:1 resonance are ultimately absorbed by the inner disk and are not accreted by the Moon because the resonance prevents close encounters between the moonlets and the Moon, while at the same time increasing moonlet eccentricities to high values that lead to close passes by the Earth and tidal disruption.  However in cases where multiple moonlets are captured into the 2:1, mutual moonlet interactions on occasion scatter objects out of resonance and allow them to be accreted by the Moon.

Thus due first to the effects of inward scattered moonlets that are lost to the inner disk, and then to moonlet capture into the 2:1 resonance with the Moon, the efficiency of accretion onto the Moon decreases substantially in the $t \approx 120$ to $t \approx \unit{200}{year}$ period, while the Moon's semi-major axis increases substantially.   This period is also associated with a rapid increase in the total mass of particles that are absorbed by the inner disk (Figure \ref{fig_sma_largest_body}, black line).
   
\begin{figure}[!h]
\begin{center}
\includegraphics[width=8cm]{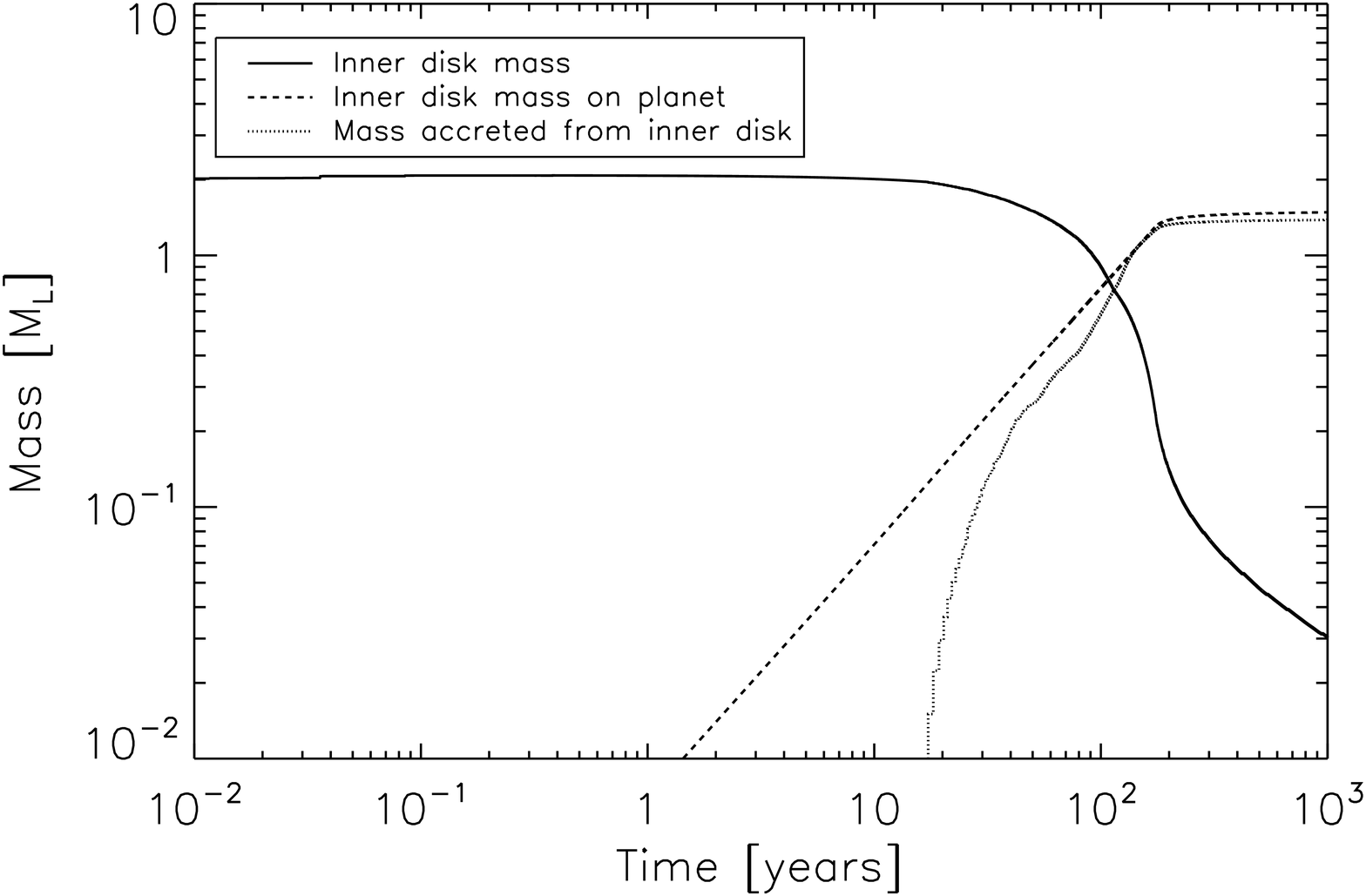}
\end{center}
\caption{Evolution of the mass of the Roche-interior disk (solid line), of the mass from the inner disk fallen onto the planet (dashed line), and of the mass of new objects accreted from the inner disk (dotted line), for Run 34.}
\label{fig_disk_mass_evolution}
\end{figure}

By $t \sim \unit{200}{years}$, the inner disk mass in Run 34 has decreased to $M_d \sim 0.2M_\leftmoon$, and the disk viscosity transitions to the Ward-Cameron viscosity.  The viscosity then decreases rapidly as the disk dissipates (since $\nu_{WC} \propto {\sigma}^2$), causing the production rate of spawned moonlets to slow dramatically (see Figure \ref{fig_disk_mass_evolution}).

Figure \ref{fig_disk_mass_evolution} shows the evolution of the Roche-interior disk mass (solid line), of the mass falling onto the planet from the Roche-interior disk (dashed lined), and of the mass of new objects accreted from the inner disk (dotted line), for Run 34. New objects are accreted at the disk's outer edge only after $\sim \unit{20}{years}$ (Figure \ref{fig_disk_mass_evolution}, dotted line). At that point, the disk has lost $\sim 0.04~M_{\leftmoon}$, that is $\sim 2\%$ of its initial mass (Figure \ref{fig_disk_mass_evolution}, dashed line). At $\sim \unit{200}{years}$, the Roche-interior disk is almost depleted in mass, and no significant mass is accreted after that point (Figure \ref{fig_disk_mass_evolution}, solid and dotted lines).

\subsubsection{Other possible outcomes}
The aforementioned dynamical steps are present in 28 of the 46 runs, but other outcomes are possible. During the first phase of accretion, interactions between initial outer bodies can lead to the Moon forming farther away, around $\sim 5R_\Earth$ or beyond. In such cases, the Moon's 2:1 resonance is far enough from the disk that newly spawned moonlets are immediately captured in the resonance. Subsequent interactions between moonlets can then result in some of them being ejected from the resonance and colliding with the Moon, just like in the general mechanism described above, or it can lead to the growth of a secondary object inside the resonance. In such cases the Moon can be driven out to $\sim 8 R_\Earth$, at which point the secondary body's semi-major axis is $\sim 4.6R_\Earth$ where it does not interact with the inner disk anymore. The interior body  can sometimes get even more massive than the body resulting from accretion of the initial outer particles (e.g. Runs 11 and 16).

For Runs 1 and 2 that do not include an initial outer disk, the outcome is similar to those described above. The first body accreted at the Roche limit confines the inner disk inside the Roche limit. As the disk viscously spreads outward, the body recoils from the disk and a second body is spawned at the Roche limit after $\sim \unit{1}{year}$. The disk is once again confined and the second body recoils until it finally merges with the first one. This process continues until the Moon gets far enough from the disk to start capturing bodies in resonance after several tens of years. At that point, interactions between particles leads to the accretion of a second large object that moves in resonance with the first one. In Run 1 the resonant configuration remains stable, while in Run 2 it eventually goes unstable due to interactions with other objects, resulting in a merger of the two largest objects.

\subsection{Simulation results}
\subsubsection{Properties of the final Moon}
\begin{figure}[!h]
\begin{center}
\includegraphics[width=8cm]{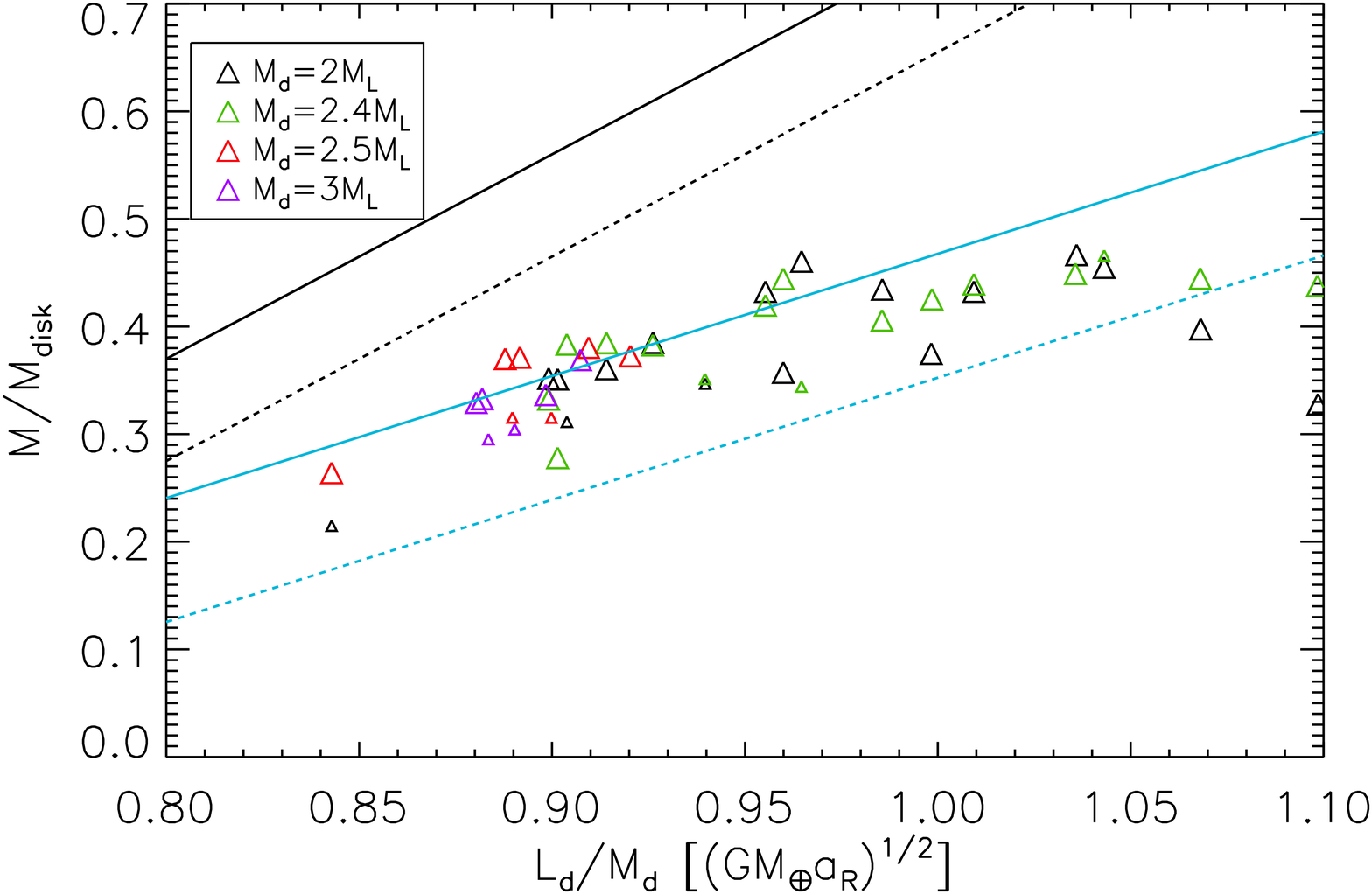}
\end{center}
\caption{Ratio of the mass of the largest body at $t = \unit{1000}{years}$ to the initial disk mass, as a function of the disk's initial specific angular momentum. Black, green, red and purple triangles correspond to Runs with a total disk mass of 2, 2.4, 2.5 and $3M_{\leftmoon}$, respectively. Small symbols are cases where the mass of the second largest body is at least $20\%$ that of the largest one. In those cases we added the mass of the two bodies, since tidal evolution could cause them to merge later on \citep{canup99}. The black solid and dashed lines correspond to equation (\ref{equ_MoonMassRelation}) with $M_\infty = 0$ and $M_\infty = 0.05 M_d$, respectively. Blue lines are the analytical estimates from equation (\ref{equ_analytical_estimate_hybrid}).}
\label{fig_Nbodyandfluid_results}
\end{figure}

The mass, angular momentum and mass fraction of inner disk material of the largest body from each of the hybrid simulations are shown in Table \ref{table_Nbodyandfluid_results}. The mass of the largest body at $t = \unit{1000}{years}$, versus the disk's initial specific angular momentum, is represented in Figure \ref{fig_Nbodyandfluid_results}. Results show a range of outcomes, with an average Moon mass of $\langle M\rangle \sim 0.81\pm 0.21~M_{\leftmoon}$, semi-major axis $\langle a\rangle = 2.15\pm 0.27 a_R$, eccentricity $\langle e\rangle = 0.042\pm 0.093$, and fraction of inner disk material $35\pm 30\%$. Accretion efficiency, defined as $M/M_d$, is somewhat lower than in pure N-body simulations, as the fraction of the disk accreted varies from $\sim20\%$ to 50\%, with the rest being either ejected from the system or lost onto the planet. As in \citet{ida97} and \citet{kokubo00}, we find that $M/M_d$ increases with the initial specific angular momentum of the disk.

\begin{deluxetable}{c c c c c c c c c c c c c c}
\tabletypesize{\footnotesize}
\tablewidth{0pt}
\tablecolumns{14}
\tablecaption{Hybrid simulations results. \label{table_Nbodyandfluid_results}}
\tablehead{ & \colhead{$a$ } & \colhead{$M$} &  &  & \colhead{$a_2$} & \colhead{$M_2$} &  & \colhead{$M_{orb}$} & \colhead{$M_{\infty}$} & \colhead{$M_{cap}$} & \colhead{$L$} & \colhead{$L_{orb}$} & \colhead{$L_{\infty}$}\\
\colhead{Run} & \colhead{$\left(a_R\right)$} & \colhead{$\left(M_\leftmoon\right)$} & \colhead{$e$} & \colhead{$f$} & \colhead{$\left(a_R\right)$}  & \colhead{$\left(M_\leftmoon\right)$} & \colhead{$f_2$} & \colhead{$\left(M_\leftmoon\right)$} & \colhead{$\left(M_\leftmoon\right)$} & \colhead{$\left(M_\leftmoon\right)$} & \colhead{$\left(L_{EM}\right)$} & \colhead{$\left(L_{EM}\right)$} & \colhead{$\left(L_{EM}\right)$}}
\startdata
1  & 2.43 & 0.231 & 0.370 & 100\% & 3.88 & 0.198 & 100\% & 0.430 & 0.027 & 0.330 & 0.060 & 0.130 & 0.011\\
2  & 2.30 & 0.660 & 0.016 & 100\% & 1.39 & 0.004 & 100\% & 0.663 & 0.011 & 0.483 & 0.180 & 0.181 & 0.004\\
\hline
3  & 1.83 & 0.865 & $< 10^{-3}$ & 3.6\%  & 1.13 & 0.001 & 100\%   & 0.866 & 0.017 & 0.409 & 0.211 & 0.211 & 0.006\\
4  & 2.44 & 0.714 & 0.004 & 10.4\% & 1.17 & 0.001 & 100\%   & 0.716 & 0.043 & 0.473 & 0.201 & 0.201 & 0.015\\
5  & 1.87 & 0.921 & 0.001 & 4.4\%  & 1.16 & 0.001 & 100\%   & 0.922 & 0.012 & 0.472 & 0.227 & 0.227 & 0.004\\
6  & 1.97 & 1.007 & 0.001 & 5.8\%  & 1.22 & 0.001 & 100\%   & 1.008 & 0.026 & 0.625 & 0.255 & 0.255 & 0.009\\
7  & 1.99 & 1.066 & 0.000 & 11.5\% & 1.22 & 0.002 & 100\%   & 1.068 & 0.025 & 0.745 & 0.271 & 0.272 & 0.008\\
8  & 2.07 & 0.682 & 0.142 & 76.4\% & 8.46 & 0.144 & 0.8\%  & 0.826 & 0.035 & 0.753 & 0.175 & 0.244 & 0.012\\
9  & 2.02 & 0.703 & 0.001 & 43.8\% & 1.24 & 0.002 & 100\%   & 0.705 & 0.034 & 0.549 & 0.180 & 0.180 & 0.012\\
10 & 2.05 & 0.702 & 0.001 & 44.5\% & 1.29 & 0.002 & 100\%   & 0.704 & 0.020 & 0.558 & 0.181 & 0.182 & 0.007\\
11 & 2.72 & 0.412 & 0.026 & 15.2\% & 1.71 & 0.211 & 100\%   & 0.623 & 0.027 & 0.321 & 0.122 & 0.171 & 0.011\\
12 & 2.20 & 0.798 & 0.001 & 30.6\% & 1.33 & 0.003 & 100\%   & 0.801 & 0.015 & 0.501 & 0.213 & 0.214 & 0.005\\
13 & 2.18 & 0.667 & 0.007 & 31.2\% & 1.36 & 0.004 & 100\%   & 0.671 & 0.127 & 0.563 & 0.177 & 0.178 & 0.041\\
14 & 1.94 & 0.920 & 0.002 & 38.1\% & 1.20 & 0.001 & 100\%   & 0.921 & 0.022 & 0.634 & 0.231 & 0.231 & 0.008\\
15 & 1.94 & 0.925 & 0.001 & 51\% & 1.21 & 0.001 & 100\%   & 0.926 & 0.030 & 0.675 & 0.232 & 0.232 & 0.011\\
16 & 1.90 & 0.397 & 0.314 & 100\%   & 3.09 & 0.392 & 19.8\% & 0.789 & 0.021 & 0.422 & 0.093 & 0.217 & 0.008\\
17 & 2.02 & 0.928 & 0.001 & 56.3\% & 1.22 & 0.001 & 100\%   & 0.930 & 0.019 & 0.689 & 0.238 & 0.238 & 0.007\\
18 & 2.09 & 0.988 & 0.002 & 52.1\% & 1.29 & 0.002 & 100\%   & 0.989 & 0.047 & 0.725 & 0.257 & 0.258 & 0.016\\
19 & 2.19 & 0.998 & 0.004 & 53.8\% & 1.35 & 0.003 & 100\%   & 1.002 & 0.028 & 0.734 & 0.266 & 0.267 & 0.010\\
20 & 1.84 & 0.532 & 0.308 & 100\%   & 2.96 & 0.354 & 27.6\% & 0.886 & 0.046 & 0.703 & 0.124 & 0.230 & 0.017\\
\hline
21 & 2.23 & 0.869 & 0.001 & 8.4\%  & 1.23 & 0.001 & 100\%   & 0.870 & 0.022 & 0.457 & 0.234 & 0.234 & 0.008\\
22 & 2.19 & 0.865 & 0.001 & 8.6\%  & 1.41 & 0.003 & 100\%   & 0.868 & 0.040 & 0.595 & 0.231 & 0.232 & 0.015\\
23 & 2.20 & 0.933 & 0.004 & 7.2\%  & 1.34 & 0.002 & 100\%   & 0.935 & 0.069 & 0.546 & 0.250 & 0.250 & 0.024\\
24 & 2.01 & 0.974 & $< 10^{-3}$ & 6.9\%  & 1.25 & 0.002 & 100\%   & 0.975 & 0.118 & 0.633 & 0.249 & 0.249 & 0.040\\
25 & 2.01 & 1.054 & 0.001 & 6.6\%  & 1.25 & 0.002 & 100\%   & 1.056 & 0.113 & 0.697 & 0.270 & 0.270 & 0.041\\
26 & 2.02 & 1.078 & 0.010 & 7.6\%  & 4.54 & 0.015 & 0\%     & 1.094 & 0.107 & 0.709 & 0.276 & 0.282 & 0.038\\
27 & 2.08 & 0.721 & 0.004 & 37.6\% & 1.31 & 0.002 & 100\%   & 0.723 & 0.021 & 0.575 & 0.188 & 0.188 & 0.008\\
28 & 2.01 & 0.770 & 0.001 & 44.9\% & 1.23 & 0.002 & 100\%   & 0.772 & 0.036 & 0.592 & 0.197 & 0.197 & 0.013\\
29 & 2.78 & 0.478 & 0.059 & 17.5\% & 1.73 & 0.216 & 100\%   & 0.694 & 0.022 & 0.220 & 0.143 & 0.192 & 0.009\\
30 & 1.98 & 0.923 & 0.001 & 45.9\% & 1.22 & 0.002 & 100\%   & 0.925 & 0.044 & 0.664 & 0.234 & 0.234 & 0.015\\
31 & 1.94 & 0.919 & 0.001 & 40.2\% & 1.22 & 0.001 & 100\%   & 0.920 & 0.055 & 0.653 & 0.231 & 0.231 & 0.019\\
32 & 2.96 & 0.559 & 0.126 & 15.6\% & 1.86 & 0.284 & 100\%   & 0.843 & 0.021 & 0.290 & 0.172 & 0.235 & 0.008\\
33 & 1.92 & 0.404 & 0.316 & 100\%   & 3.04 & 0.384 & 15.9\% & 0.788 & 0.037 & 0.420 & 0.096 & 0.216 & 0.015\\
34 & 1.97 & 0.951 & 0.000 & 56.6\% & 1.22 & 0.001 & 100\%   & 0.952 & 0.034 & 0.691 & 0.241 & 0.241 & 0.012\\
35 & 1.99 & 0.932 & 0.000 & 50\% & 1.22 & 0.002 & 100\%   & 0.933 & 0.050 & 0.684 & 0.237 & 0.237 & 0.018\\
36 & 2.86 & 0.517 & 0.062 & 14.3\% & 1.75 & 0.397 & 100\%   & 0.913 & 0.056 & 0.441 & 0.157 & 0.249 & 0.019\\
37 & 2.13 & 1.010 & 0.002 & 56.1\% & 1.33 & 0.003 & 100\%   & 1.013 & 0.077 & 0.735 & 0.265 & 0.266 & 0.027\\
38 & 2.05 & 1.107 & 0.001 & 58.2\% & 1.27 & 0.002 & 100\%   & 1.109 & 0.038 & 0.773 & 0.286 & 0.286 & 0.014\\
\hline
39 & 2.05 & 0.795 & 0.002 & 19.3\% & 4.27 & 0.012 & 0.2\%  & 0.809 & 0.215 & 0.657 & 0.205 & 0.210 & 0.077\\
40 & 2.33 & 1.067 & 0.002 & 7\%  & 1.10 & $< 10^{-3}$ & 100\%   & 1.068 & 0.097 & 0.536 & 0.293 & 0.294 & 0.038\\
41 & 2.63 & 0.750 & 0.050 & 13\% & 1.64 & 0.058 & 100\%   & 0.808 & 0.015 & 0.297 & 0.219 & 0.231 & 0.006\\
42 & 2.02 & 0.910 & 0.007 & 12.1\% & 4.14 & 0.038 & 0.3\%  & 0.949 & 0.080 & 0.624 & 0.233 & 0.247 & 0.027\\
43 & 2.00 & 0.656 & 0.014 & 18.2\% & 9.02 & 0.012 & 0\%  & 0.669 & 0.292 & 0.828 & 0.167 & 0.173 & 0.106\\
44 & 2.29 & 1.021 & 0.001 & 8.8\%  & 1.45 & 0.003 & 100\%   & 1.024 & 0.060 & 0.652 & 0.278 & 0.279 & 0.021\\
45 & 1.93 & 0.897 & 0.070 & 10.7\% & 4.45 & 0.221 & 0.2\%  & 1.119 & 0.058 & 0.847 & 0.224 & 0.304 & 0.020\\ 
46 & 2.51 & 1.051 & 0.002 & 9.3\%  & 1.19 & 0.001 & 100\%   & 1.052 & 0.122 & 0.368 & 0.300 & 0.300 & 0.048\\
\enddata
\tablecomments{$a$, $e$, $M$, $f$ and $L$ are the semi-major axis, eccentricity, mass, mass fraction of inner disk material, and angular momentum of the largest Moon at the end of the simulation ($t = \unit{1000}{years}$). $a_2$, $M_2$ and $f_2$ are the semi-major axis, mass, and mass fraction of inner disk material of the second largest body. $M_{orb}$ and $L_{orb}$ are the mass and angular momentum of all orbiting bodies at $t = \unit{1000}{years}$. $M_{\infty}$ and $L_{\infty}$ are the mass and angular momentum of ejected particles. $M_{cap}$ is the total mass of bodies that where tidally disrupted and captured in the inner disk. Units of mass, distance and angular momentum are the present Lunar mass $M_{\leftmoon}$, Roche limit for silicates $a_R\approx 2.9R_{\Earth}$, and angular momentum of the Earth-Moon system $\left(L_{EM} = \unit{3.5\times10^{41}}{\gram\usk\centi\squaremetre\usk\reciprocal\second}\right)$}
\end{deluxetable}

\begin{figure}[!h]
\begin{center}
\plottwo{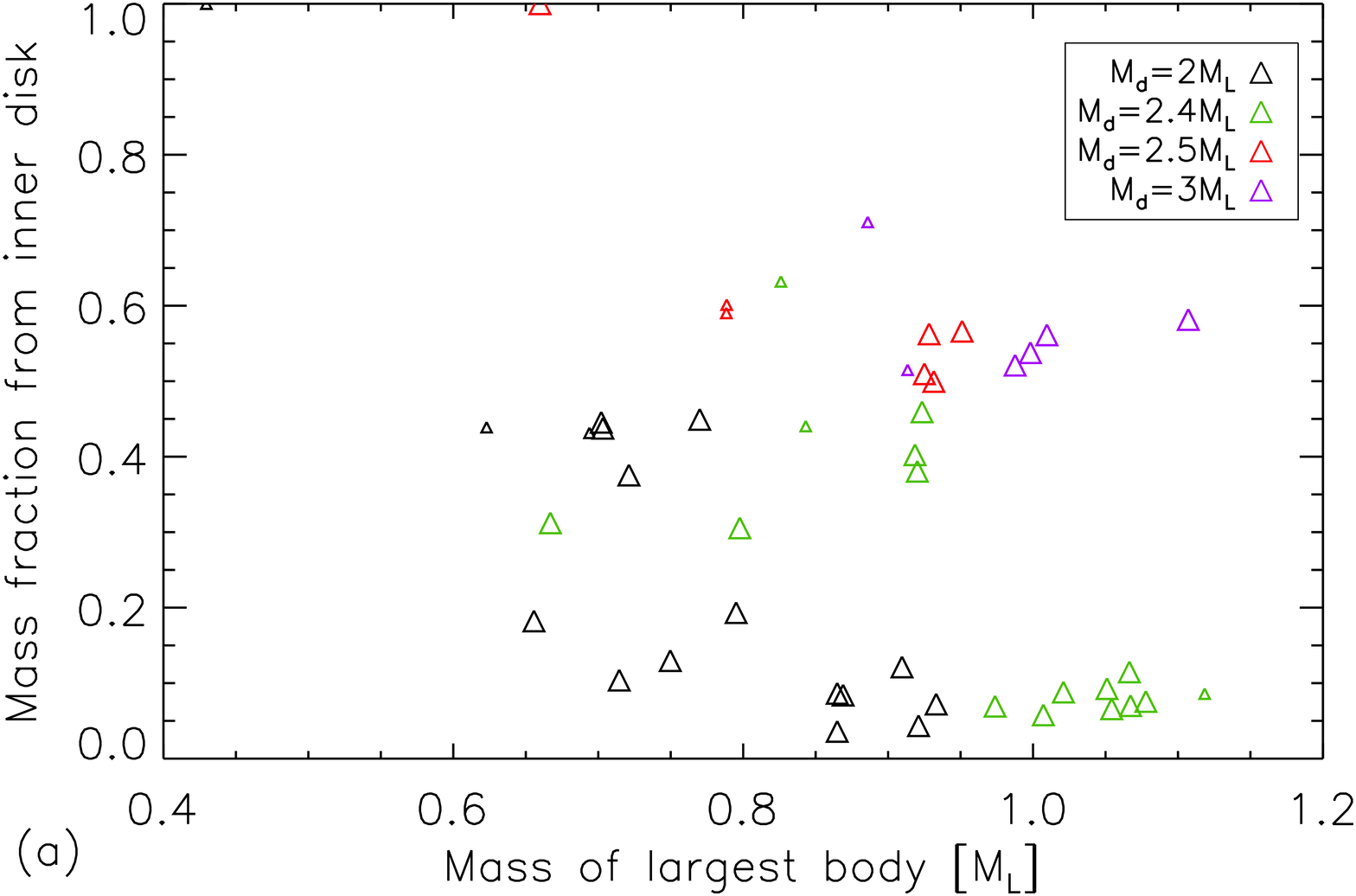}{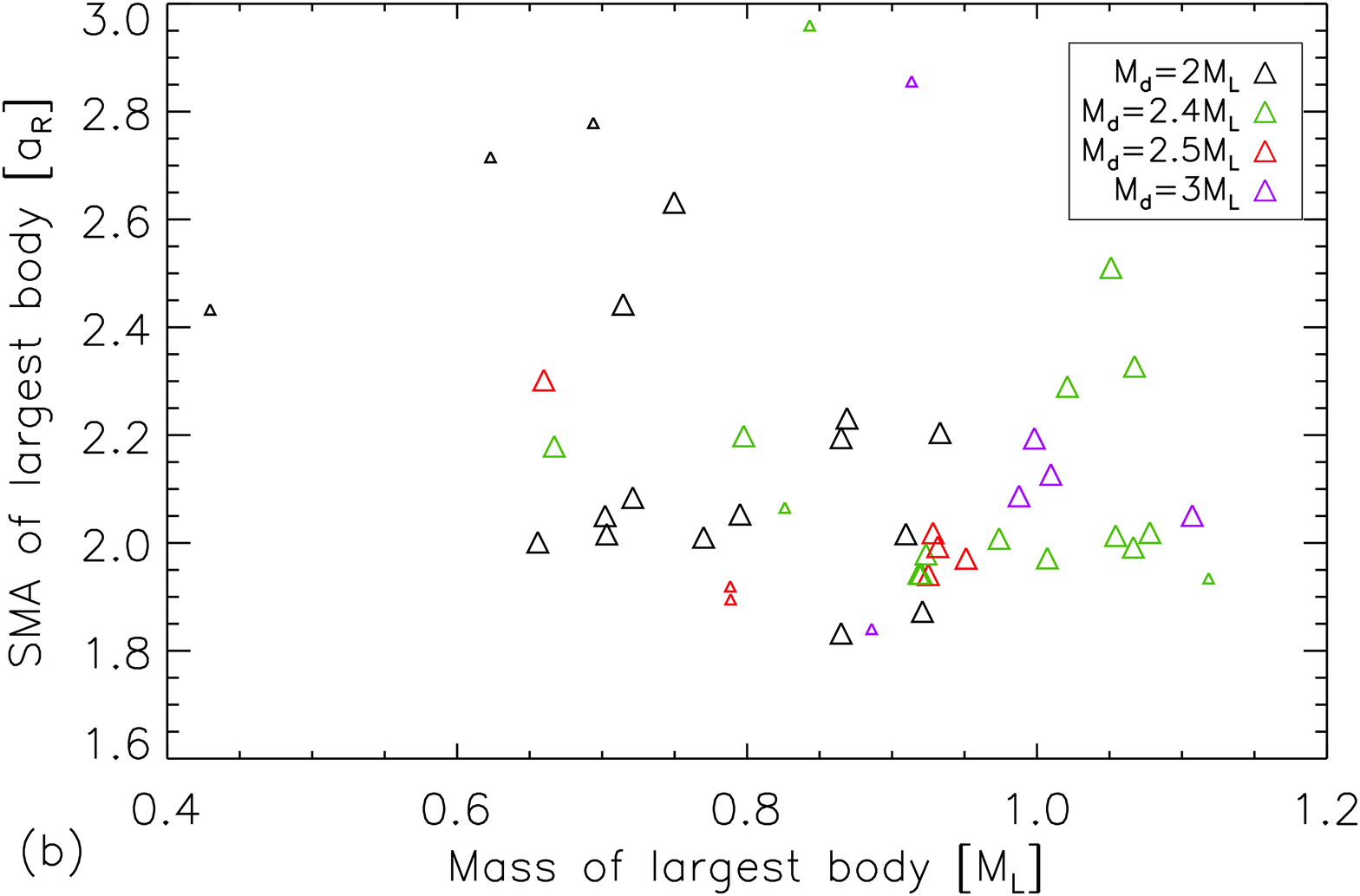}
\epsscale{0.45}
\plotone{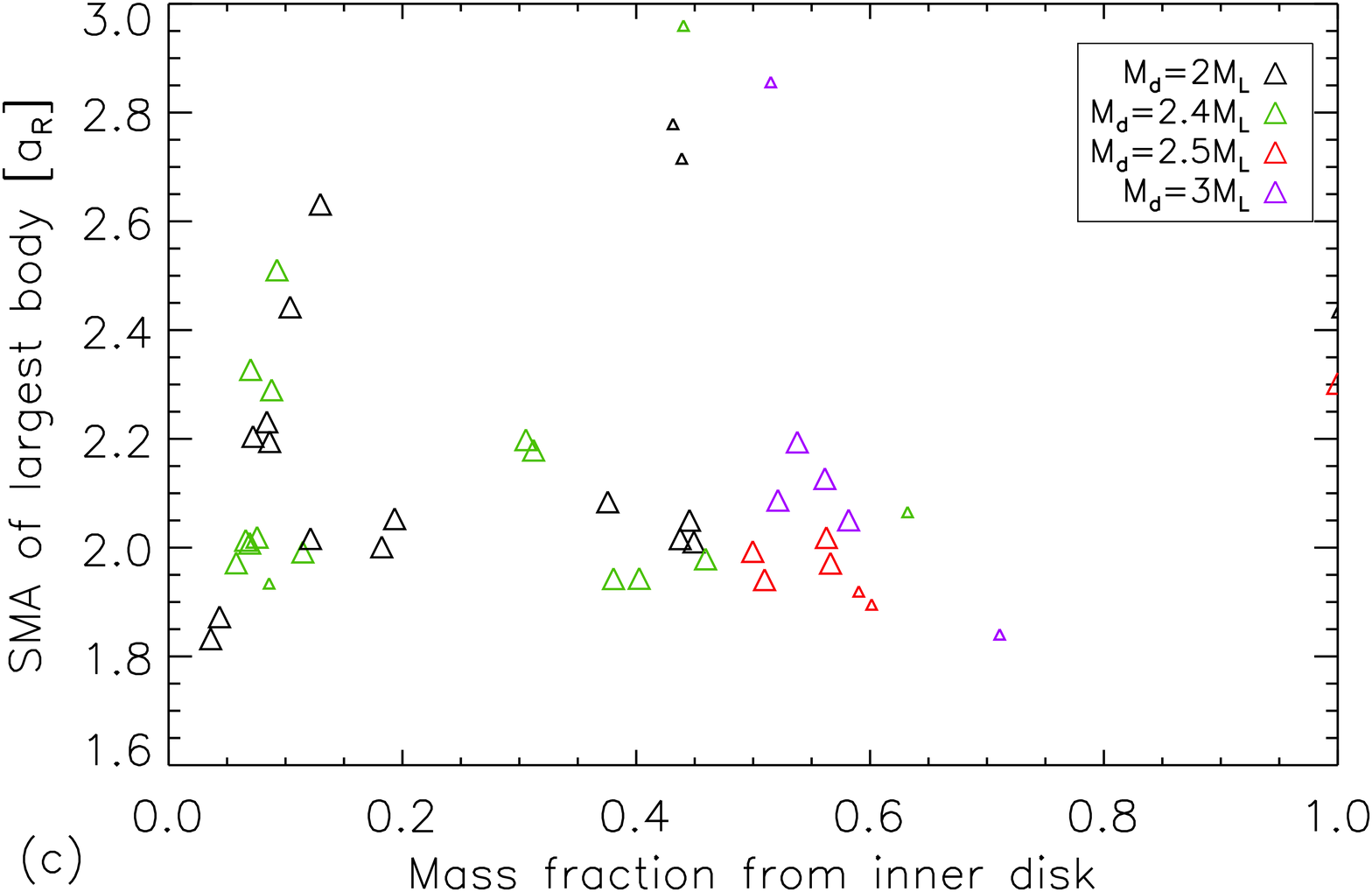}
\end{center}
\caption{a) Fraction of the mass of the largest body composed of Roche-interior disk material, against mass of the largest body. b) Semi-major axis of the largest body against its mass. c) Semi-major axis of the largest body against its mass fraction of inner disk material. Black, green, red and purple triangles correspond to Runs with a total disk mass of 2, 2.4, 2.5 and $3M_{\leftmoon}$, respectively. Small symbols are cases where the mass of the second largest body is at least $20\%$ that of the largest one. In those cases we added the mass of the two bodies. The two points at 100\% inner disk material correspond to Runs 1 and 2 that include only a Roche-interior disk initially.}
\label{fig_mass_fraction_vs_mass}
\end{figure}

Figure \ref{fig_mass_fraction_vs_mass}a shows the mass of the largest body at $t = \unit{1000}{years}$ versus the fraction of its mass that is composed of particles accreted from the inner disk. Figure \ref{fig_mass_fraction_vs_mass}b shows the semi-major axis of the largest body against its mass. Figure \ref{fig_mass_fraction_vs_mass}c shows the semi-major axis of the largest body against its mass fraction of inner disk material. For the less massive initial disks, final moons with a mass $\ge 0.8 M_\leftmoon$ generally contain less than 20\% of their mass in material originating from the inner disk. This is because the Roche-interior disk is more strongly confined when a larger object is formed by accretion of the initial outer bodies, and while the disk is confined it loses a significant mass onto the planet. Starting with initially more massive disks (red and purple points, corresponding to total disk masses of 2.5 and 3$M_\leftmoon$) increases this fraction to $\sim 60\%$. However lunar-forming impact simulations have not generally produced such massive disks for cases in which the impact angular momentum is comparable to $L_{EM}$.

\subsubsection{Analytical estimate}{\label{section_analytical_estimate}}
While analytical estimates from equation (\ref{equ_MoonMassRelation}) are in good agreement with pure N-body simulations (Figure \ref{fig_Nbody_results}), this is no longer the case for simulations with a Roche-interior fluid disk (Figure \ref{fig_Nbodyandfluid_results}, black lines). To derive formula (\ref{equ_MoonMassRelation}), \citet{ida97} assumed that the Moon formed at $1.3 a_R$, while in our hybrid simulations $\langle a \rangle \sim 2.15 a_R$. To revise this formula, we redo the calculation in \citet{ida97} and consider conservation of the disk's angular momentum, which gives
\begin{equation}
 L_d = M_1\sqrt{GM_{\Earth}\left(1-e_1^2\right)a_1} + M_{pl}\sqrt{GM_{\Earth}\left(1-e_2^2\right)a_2} + M_\infty\sqrt{GM_{\Earth}\left(1-e_3^2\right)a_3},
\label{equ_disk_angular_momentum_conservation}
\end{equation}
where $M_1$, $a_1$ and $e_1$ are the mass, semi-major axis and eccentricity of the Moon, $M_{pl}=M_d-M_1-M_\infty$ is the mass scattered onto the planet (solid bodies and through the inner edge of the inner disk), with semi-major axis and eccentricity $a_2$ and $e_2$, and $M_\infty$ is the mass of ejected bodies, with semi-major axis and eccentricity $a_3$ and $e_3$. Here we assume that all material initially in the disk is either accreted by the Earth, accreted into a Moon at $a=a_1$, or scattered onto escaping trajectories. Average values from Table \ref{table_Nbodyandfluid_results} give $e_1 \approx 0.04$ so that $(1-e_1^2) \approx 1$. Since most of the material accreted by the Earth comes from the inner disk (due to tidal disruption, generally no particles collide with the Earth), we can set $e_2 \approx 0$ and $a_2 = R_{\Earth}$. Finally, assuming that the Moon scatters escaping material on nearly hyperbolic orbits, we can set $(1-e_3)a_3 \sim a_1$ and $(1+e_3)\sim 2$. Equation
 (\ref{equ_disk_angular_momentum_conservation}) then becomes
\begin{equation}
 \frac{L_d}{M_d\sqrt{GM_{\Earth}a_R}} = \frac{M_1}{M_d}\sqrt{\frac{a_1}{a_R}} + 
                                                 \frac{\left(M_d-M_1-M_\infty\right))}{M_d}\sqrt{\frac{R_{\Earth}}{a_R}} + 
                                                 \frac{M_{\infty}}{M_d}\sqrt{\frac{2a_1}{a_R}}.
\end{equation}

Then, using $\sqrt{a_R/R_\Earth}=1.7$ we get
\begin{equation}
\frac{M_1}{M_d} = \frac{1.7}{\sqrt{a_1/R_\Earth}-1}\frac{L_d}{M_d\sqrt{GM_{\Earth}a_R}} - \frac{1}{\sqrt{a_1/R_\Earth}-1} -                    \frac{M_{\infty}}{M_d}\frac{\sqrt{2a_1/a_R}-1}{\sqrt{a_1/a_R}-1}.
\label{equ_analytical_estimate_general2}
\end{equation}

For hybrid simulations, assuming the Moon forms at $a_1 \approx 2.15 a_R$ so that $a_1/R_\Earth \approx 6.2$, we get
\begin{equation}
\frac{M_1}{M_d} = 1.14\frac{L_d}{M_d\sqrt{GM_{\Earth}a_R}} - 0.67 - 2.3\frac{M_{\infty}}{M_d}.
\label{equ_analytical_estimate_hybrid}
\end{equation}
This equation is plotted on Figure \ref{fig_Nbodyandfluid_results} as the blue solid and dashed lines, for $M_\infty=0$ and $M_\infty=0.05M_d$ respectively, which show good agreement with the results from our simulations.

\section{Discussion}
\subsection{Summary}
We have developed a new numerical model to study the formation of the Earth's Moon from an impact generated disk: an N-body symplectic integrator coupled to a simple model for a fluid Roche-interior disk. Our model includes: (1) viscous spreading of the Roche interior disk, using either an instability-driven viscosity, or a radiation-limited viscosity, (2) accretion of moonlets when the inner disk's outer edge reaches the Roche limit, (3) tidal accretion criteria to treat collisions between orbiting bodies, (4) disk-satellite interactions at $0^{th}$ order Lindblad resonances, (5) spawning of new moonlets as the inner disk spreads past the Roche limit, and (6) tidal disruption of objects scattered close to the planet.

Our initial setup consists of a fluid disk extending from the Earth's surface to the Roche limit at $2.9R_{\Earth}$, and individual particles beyond. We find that the Moon accretes in 3 consecutive phases, accreting first from the bodies initially present outside the Roche limit, which confine the inner disk within the Roche limit. The inner disk slowly viscously spreads back out to the Roche limit, pushing along outer bodies via resonant interactions. After several tens of years, the disk spreads beyond the Roche limit, and starts producing new objects that continue the growth of the Moon, until the inner disk is depleted in mass after several hundreds of years. For initial disk masses in the range impact simulations typically predict, a moon with a mass of 0.6 to 1.1 $M_\leftmoon$ is produced, with a mass fraction of Roche-interior material of 5 to 65\%, accreted only during the last stage of the Moon's accretion. Increasing the initial total mass of the disk can produced large moons containing up to 60\% inner disk material, although it is not clear yet if such disks could be produced by appropriate giant impacts.

Most of our simulations produce a single Moon outside the Roche limit, similar to prior pure N-body models (Ida et al. 1997; Kokubo et al. 2000).  However there are several key differences.  First, consideration of a fluid inner disk leads to a lengthening of the Moon's total accretion timescale to $\sim \unit{10^2}{years}$, vs. a timescale of $< \unit{1}{year}$ predicted by pure N-body models.  Material that rapidly accretes outside the Roche limit resonantly confines the inner disk, which delays the accretion of the inner disk material until the disk can viscously spread back out to the Roche limit.  For a fluid disk with a thermally regulated viscosity \citep{thompson88}, the latter typically requires $\geq \unit{50}{years}$. The inner disk material is then preferentially accreted during the last stages of the Moon's growth.  In contrast, in pure N-body models the viscosity of the inner disk is large and the disk spreads very quickly (in $< \unit{1}{year}$), so that inner and outer disk material is accreted more-or-less simultaneously by the growing Moon.   

The prolonged period of interaction between the fluid inner disk (and moonlets spawned from it) and the outer Moon also leads to a substantially larger semi-major axis for the final Moon, with $\langle a\rangle \approx 2.15 a_R$ vs. $\langle a\rangle \approx 1.3a_R$ in the pure N-body simulations. A larger initial semi-major axis for the Moon in turn implies a somewhat lower overall accretion efficiency for a given initial disk mass and angular momentum, with our hybrid simulations finding that only 20 to 50\% of the initial total disk mass is ultimately incorporated into the Moon.  This suggests that an initial disk mass $\ge 2M_\leftmoon$ is required to produce a lunar mass Moon, which is somewhat larger than that produced to date by most impact simulations (e.g., Canup et al. 2012) that produce a planet-disk system whose angular momentum is comparable to that in the current Earth and Moon.

\subsection{Relation to equilibration}
A key constraint on prior impact simulations is the present angular momentum of the Earth-Moon system, which constrains the impactor size, the impact angle, and the relative velocity. For an impact angular momentum comparable to $L_{EM}$, the outcome of the impact is a circumterrestrial disk composed primarily of impactor material \citep[e.g.][]{benz89,canup04a}. If the Moon accreted from such a disk, it would then have a composition close to that of the impactor. 

The Earth-Moon system shows striking compositional similarities, in particular regarding oxygen isotopes \citep{wiechert01}. However, the distribution of this element in the early solar system was very heterogeneous \citep{clayton93}. In addition, the scale of radial mixing found in terrestrial accretion simulations \citep{chambers01} implies that the impactor would have had a substantially different composition from that of the Earth \citep{pahlevan07}, which contradicts the observed similarities.

It has been suggested that mixing could occur between the disk's atmosphere and that of the Earth, leading to the equilibration of disk-planet compositions in $10^2 - \unit{10^3}{years}$ \citep{pahlevan07}. This is much longer than the accretion timescales predicted by N-body simulations. However, in our model, the slow spreading of the disk delays the final accretion of the Moon by several hundreds of years, which could be compatible with estimated equilibration timescales.

The 3-step accretion mechanism revealed in our simulations implies that only material accreted during the final stage would have been processed through the Roche-interior disk. Earth-like material could then naturally end up in the outer parts of the Moon, although mixing in the lunar interior would need to be taken into consideration.

We can however adopt an idealized model for the Moon where initial outer disk bodies accrete into a \textit{core} with radius $R_1$, and material processed in the inner disk piles up later to increase the radius to $R_2$. Noting $f$ the mass fraction of inner disk material, we can express $R_1$ and $R_2$ as \begin{equation}
 R_1 = \left[\frac{3M\left(1-f\right)}{4\pi\rho}\right]^{1/3}
\end{equation}
and
\begin{equation}
 R_2 = \left[\frac{3M}{4\pi\rho}\right]^{1/3}
\end{equation}

For $M=1~M_{\leftmoon}$, $f=50\%$ and $\rho=\unit{3500}{\kilo\gram\usk\rpcubic\metre}$, we get $R_1\approx \unit{1358}{\km}$ and $R_2\approx \unit{1711}{\km}$. In the limit that no mixing occurs between the early and late-accreted material, the Earth-like material would represent a $R_2-R_1\approx \unit{350}{\km}$-deep outer layer on the Moon. Whether this would be sufficient to explain the identical composition between Earth and the lunar samples is not clear.

\subsection{Model limitations}{\label{section_model_limitations}}
\subsubsection{Uniform surface density inner disk}
In our model we assume the disk maintains a uniform surface density profile. Numerical simulations of the viscous evolution of self-gravitating dense planetary rings show that the disk evolves with a density peak inward and lower densities in the outer regions, regardless of the disk's initial profile \citep{salmon10}. In the instability-driven regime, this would increase (decrease) the viscosity close to the planet (at the Roche limit). The opposite would happen in the radiation-limited regime (see equations (\ref{equ_viscosity_WC}) and (\ref{equ_viscosity_TS})). A higher viscosity and/or surface density close to the Roche limit would decrease the ability of exterior moonlets to confine the inner disk, since the balance between the viscous and resonant torque would be more difficult to achieve. As a result, more material may be brought viscously through the Roche limit, thus possibly improving the accretion efficency, resulting in a larger moon formed for a given disk mass, compared to the slab model. However, since the mass necessary to confine the inner disk with the present model is so much smaller than a lunar mass, confinement of the inner disk and the associated phase (2) of the accretion process seem inevitable when forming a lunar-mass Moon (see Section \ref{section_three_stage_accretion}). Simulation of the radial, as well as temporal, evolution of the inner disk is certainly possible \citep[e.g.][]{charnoz10,salmon10}, although more computationally intensive, and such modeling is planned in our future work.

Another simplification adopted for the viscous spreading of the fluid disk is the computation of the mass fluxes at the disk's inner and outer edges. Both of these fluxes are estimated using the viscosity at the Roche limit. However, the viscosity varies with distance as $\nu_{WC} \propto r^{9/2}$ or $\nu_{TS} \propto r^3$. Thus we overestimate the rate of mass loss onto the planet, and we expect that future models that include the variation of viscosity with distance may increase the disk lifetime. As a consequence, we can expect more material to be delivered through the Roche limit, thus increasing the fraction of potentially equilibrated material incorporated in the final Moon.

\subsubsection{Co-evolution of liquid and gas phases}
Our inner disk model assumes that both the vapor and condensed phases viscously evolve as a single unit. This is motivated by the \cite{thompson88} disk model in which the liquid and vapor phases remain vertically well-mixed. Recently \citet{ward12} has developed a generalized description of vertical disk structures appropriate for a two-phase silicate protolunar disk. He identifies alternative disk solutions that involve a stratified disk, in which the condensates settle to the disk mid-plane and are surrounded by a gravitationally stable vapor atmosphere. The mid-plane layer then has a large, instability induced viscosity (per equation (\ref{equ_viscosity_WC})), while the atmospheric viscosity could be much smaller. The mid-plane layer surface density regulates itself so that the energy dissipated matches that which can be radiated from the surface of a $\sim 2000 K$ vapor disk \citep{ward12}. 

To describe such a structure will require separate tracking of the condensate and vapor layers, which we plan in future work. How might this affect the truncation of the inner disk by the outer moon(s)? Initially resonant torques will cause the outer edge of the liquid layer to contract inward relative to the outer edge of the gas disk. Gas that lies beyond the outer edge of the liquid layer may then condense (because it was the energy supplied by the underlying condensate layer that was keeping it in the vapor phase) so that there will be a condensation front that will lie outside the liquid layer's outer edge. Once gas has condensed into liquid, the liquid will be subject to resonant torques and truncated in a similar manner to that found here.

\subsubsection{Resonances}
For each satellite, we determine which of its resonances fall in the disk and the associated torque. The total torque exerted by all orbiting objects onto the disk is then applied at the disk's outer edge. As a result, the confinement of the Roche-interior disk may be too efficient. A more realistic model that applies torques at the location of each resonance in the disk \citep[as in ][]{charnoz10} could increase the ability of inner disk material to spread outward and be accreted by the Moon, bringing more potentially equilibrated material to the Moon.

Our model adopts the standard resonant torque expressions appropriate for a cold disk.  For a hot disk, the positions of the inner Lindblad resonances are shifted inward (and their torques correspondingly reduced).  A revised torque expression for the $(m:m-1)$ resonance is \citep{papaloizou07}:

\begin{equation}
\Gamma_m = \frac{\pi^2\sigma}{3\Omega\Omega_{ps}\sqrt{1+\xi^2}\left(1+4\xi^2\right)}\Psi^2,
\end{equation}
where $\Psi$ is the satellite's gravitational potential, $\xi = mh$, and $h$ is the disk's aspect ratio, with $h \approx 0.1$ for the protolunar disk\citep[see][their Table 1]{thompson88}. This expression reduces to that given by equation (\ref{equ_torque_lindblad}) for $\xi \ll 1$.

The high $m$ resonances that are closest to a satellite are the most affected, which will principally impact the initial recoil of objects close to the disk's edge. Once a satellite migrates outward away from the disk's edge, only its resonances of lower order are in the disk. Of particular importance is the 2:1 resonance --- which we argue is responsible for the initiation of phase 2 and the nature of the Moon's accretion in phase 3 --- and the torque due to this resonance is reduced by about 15\% due to thermal effects for $h = 0.1$.  We performed a test simulation using the modified torque expression above and found no substantial modifications to the overall accretion history described in Section \ref{section_accretion_dynamics}.

\subsubsection{Size of fragments}
To improve computation efficiency, we set the smallest fragment that can be spawned from the inner disk to $10^{-7}M_\Earth$, although fragments some two orders of magnitude smaller than this are predicted when the inner disk's mass decreases to $\sim10^{-1}M_\leftmoon$. This should not significantly impact the outcome of a given simulation, as such small fragments are produced when the disk is almost fully depleted in mass, so that no further growth of the moon is expected. To check the influence of this parameter, we compare results of simulations using 4 values: no limit, $10^{-9}M_\Earth$, $10^{-8}M_\Earth$, and $10^{-7}M_\Earth$. Table \ref{table_moonlet_size_influence} shows the resulting mass, semi-major axis, eccentricity, and fraction of inner disk material for the largest and second largest body, at $t = \unit{1000}{years}$, for 2 test runs with parameters similar to Run 22 (chosen arbitrarily). Although the Moon's predicted semi-major axis and eccentricity do increase somewhat as the fragment mass is reduced (due to an increased number of particles captured into resonance as their size decreases), results show that the final properties of the Moon are not strongly affected by the minimal mass set for the accretion of new moonlets.

\begin{deluxetable}{c c c c c c c c c}
\tabletypesize{\footnotesize}
\tablewidth{0pt}
\tablecolumns{9}
\tablecaption{Influence of minimal fragment size. \label{table_moonlet_size_influence}}
\tablehead{ & \colhead{$a$ } & \colhead{$M$} &  &  & \colhead{$a_2$} & \colhead{$M_2$}\\
\colhead{Fragment mass} & \colhead{$\left(a_R\right)$} & \colhead{$\left(M_\leftmoon\right)$} & \colhead{$e$} & \colhead{$f$} & \colhead{$\left(a_R\right)$}  & \colhead{$\left(M_\leftmoon\right)$} & \colhead{$e_2$} & \colhead{$f_2$}}
\startdata
\cutinhead{Test Run 1}
Unlimited & 2.82 & 0.485 & 0.099 & 14.60\% & 1.76 & 0.131 & 0.420 & 100\%\\
$10^{-9} M_\earth$ & 2.82 & 0.485 & 0.100 & 14.59\% & 1.76 & 0.131 & 0.420 & 100\%\\ 
$10^{-8} M_\earth$ & 2.78 & 0.485 & 0.071 & 14.66\% & 1.76 & 0.134 & 0.406 & 100\%\\
$10^{-7} M_\earth$ & 2.65 & 0.487 & 0.017 & 14.98\% & 1.67 & 0.141 & 0.183 & 100\%\\
\\
\cutinhead{Test Run 2}
Unlimited & 2.58 & 0.644 & 0.043 & 26.79\% & 1.61 & 0.020 & 0.557 & 100\%\\
$10^{-9} M_\earth$ & 2.58 & 0.644 & 0.043 & 26.79\% & 1.61 & 0.020 & 0.557 & 100\%\\
$10^{-8} M_\earth$ & 2.55 & 0.643 & 0.005 & 26.78\% & 1.59 & 0.021 & 0.283 & 100\%\\
$10^{-7} M_\earth$ & 2.56 & 0.640 & 0.019 & 26.39\% & 1.59 & 0.028 & 0.259 & 100\%\\
\enddata
\tablecomments{$a$, $e$, $M$, and $f$ are the semi-major axis, eccentricity, mass and mass fraction of inner disk material of the largest moon at the end of the simulation ($t = \unit{1000}{years}$). $a_2$, $e_2$, $M_2$ and $f_2$ are the semi-major axis, eccentricity, mass and mass fraction of inner disk material of the second largest body. Units of mass and distance are the present Lunar mass $M_{\leftmoon}$, and the Roche limit for silicates $a_R\approx 2.9R_{\Earth}$}
\end{deluxetable}

\subsection{Physical state of accreting material}{\label{section_physical_state}}
In the immediate aftermath of the giant impact, ejected material with equivalent circular orbits exterior to the Roche limit is predicted to have temperatures $\sim 2000\kelvin$ to $5000{\kelvin}$ \citep[e.g.][Fig. 6]{canup04a}, and to be predominantly silicate melt (with $\sim O(10\%)$ vapor by mass). If the initial outer disk surface density of melt/solid, $\sigma_S$, is high enough for gravitational instability, the disk will on a rapid, orbital timescale, fragment into clumps with radii $R \sim O(10)~{\km}$ \citep[e.g.][]{thompson88}. The subsequent timescale for solids in the outer disk to grow through binary collisions is 
\begin{equation}
\tau_{coll} \sim \frac{R\rho}{\sigma_S\Omega} 
           \sim 0.05~\text{yr} \left(\frac{R}{\unit{10^3}{\km}}\right)\left( \frac{\sigma_S}{\unit{10^6}{\gram\usk\centi\rpsquare\metre}}\right)\left(\frac{a}{a_R}\right)^{3/2},
\end{equation}
where $\rho$ is the bulk density of the solid particles. The timescale for radiative cooling from the surfaces of a vertically well-mixed disk is
\begin{equation}
 \tau_{cool} \sim \frac{\sigma_S C_P}{2\sigma_{SB}T^3} 
            \sim 1~\text{yr} \left( \frac{\sigma_S}{\unit{10^6}{\gram\usk\centi\metre\rpsquared}}\right) \left( \frac{T}{\unit{1500}{\kelvin}}\right)^{-3},
\end{equation}
where $T$ is the disk temperature, $\sigma_{SB}$ is the Stefan-Boltzman constant, and a specific heat $C_P \sim\unit{10^7}{\erg\usk\reciprocal\gram\usk\reciprocal\kelvin}$ is assumed.

Because the time for the disk to cool to temperatures below the solidus is longer than the accretion timescale in the outer disk, material initially orbiting outside the Roche limit will accrete in a hot, molten state. \citet{pritchard00} estimate that protolunar disk material orbiting between 2 and 5 Earth radii will take of order $\unit{10}{years}$ to lose memory of the high temperatures produced by the giant impact, and find that individual large objects $\left( R > \unit{100}{\km}\right)$ can retain temperatures in excess of $\unit{1000}{\kelvin}$ for $\unit{10^2}{years}$, even given conditions designed to maximize cooling (e.g., neglecting the energy of accretion itself).

Our simulations reveal two stages of accretion: an early, rapid phase in which material initially placed outside the Roche limit by the impact accretes in $\sim \unit{0.1}{years}$, and a protracted phase in which material is delivered to the outer disk on a much longer timescale of $\sim \unit{10^2}{years}$, as the Roche interior disk viscously spreads. During phase 1, accreting objects will be inevitably hot and at or above the solidus, while cooling and some solidification might occur during phase 2.

\section*{Acknowledgements}
This work has been funded by NASA's Lunar Advanced Science and Exploration Research (LASER) program and the NASA Lunar Science Institute (NLSI). We thank William Ward for valuable comments, and for the diffusion model in Appendix \ref{appendix_disk_mass} that he developed for \citet{ward00} and \citet{canup00}.

\newpage

\appendix
\begin{center}
\bf
APPENDIX\\
Details on the disk model
\end{center}

\section{Evolution of inner disk mass}{\label{appendix_disk_mass}}
We consider a simple diffusion model for a uniform surface density inner disk by calculating the effective changes in its inner and outer edges, $R$ and $r_d$, under the constraint that $\sigma$ is uniform with distance $r$. This is the same disk model developed by W. R. Ward for \citet{ward00} and \citet{canup00}.

The disk spreading timescale is $t_{visc} = \Delta R^2 / \nu$ where $\Delta R = r_d - R$ is the disk's width and $\nu$ is its viscosity. Differentiating with respect to time gives
\begin{equation}
1 = \frac{2\left( r_d - R \right)\left( \dot{r}_d - \dot{R} \right)}{\nu},
\end{equation}
where we assume a constant viscosity for simplicity. This assumption is fairly accurate as long as integration timesteps remain small, which will be the case here. This yields
\begin{equation}
\dot{r}_d  = \frac{\nu}{2\left(r_d - R\right)} + \dot{R}
\label{equrdR1}
\end{equation}
as the rate of expansion of the disk's outer edge.

The inner disk angular momentum is
\begin{align}
\nonumber L_d & = \frac{4}{5} \pi \sigma \sqrt{GM_{\Earth}}\left( r_d^{5/2} - R^{5/2}\right)\\
& = \frac{4}{5} M_d \sqrt{GM_{\Earth}}\frac{r_d^{5/2} - R^{5/2}}{r_d^2 - R^2},\label{equ_InnerDiskAngularMomentum}
\end{align}
where $M_d = \sigma\pi\left(r_d^2 - R^2\right)$ is the disk's mass. Viscous spreading yields no net torque on the disk, implying 
\begin{equation}
\frac{dL_d}{dt} = 0 = \frac{d}{dt}\left( \frac{r_d^{5/2} - R^{5/2}}{r_d^2 - R^2} \right),
\end{equation}
which can be rewritten as
\begin{equation}
\dot{R} = \frac{4x\left(x^{5/2}-1\right) - 5\left(x^2-1\right)x^{3/2}}{4\left(x^{5/2}-1\right) -5\left(x^2-1\right)}\dot{r}_d \equiv f(x)\dot{r}_d,
\label{equrdR2}
\end{equation}
where $x\equiv r_d/R$. Finally, combining equations (\ref{equrdR1}) and (\ref{equrdR2}) gives 
\begin{equation}
\left.\dot{r}_d\right|_{visc} = \frac{\nu}{2R\left(x-1\right)\left(1 - f(x)\right)}
\label{equrddot}
\end{equation}
\begin{equation}
\dot{R} = \frac{\nu f(x)}{2R\left(x-1\right)\left(1 - f(x)\right)},
\label{equRdot}
\end{equation}
as the rates of change of the disk's inner and outer edges due to viscous spreading.

We set $R = R_{\Earth}$, so that the rate of mass loss from the disk due to infall onto the planet is
\begin{equation}
\left.\frac{dM_d}{dt}\right|_{P} = 2\pi R\dot{R}\sigma=\frac{2\pi R\dot{R}M_d}{\pi R^2 \left( x^2-1 \right)},
\end{equation}
with $\dot{R}$ from equation (\ref{equRdot}). Once $r_d$ expands to the Roche limit, we consider that material that diffuses beyond $a_R$ accretes into moonlets that are added to the N-body code (see Section \ref{section_spawning}), and this results in an additional loss of mass from the inner disk at a rate 
\begin{equation}
\left.\frac{dM_d}{dt}\right|_{a_R} = 2\pi r_d\dot{r}_d\sigma=\frac{2\pi r_d\dot{r}_d M_d}{\pi R^2 \left( x^2-1 \right)}.
\end{equation}
Here $\dot{r_d} = \left.\dot{r_d}\right|_{visc} + \left.\dot{r_d}\right|_{moon}$, where the first term is from equation (\ref{equrddot}) and the second term modifies the expansion rate of the outer edge due to satellite torques, per equation (\ref{rdotmoon}) below. The total rate of change in the inner disk mass is then
\begin{equation}
\left.\frac{dM_d}{dt}\right|_{visc} = \left.\frac{dM_d}{dt}\right|_P + \left.\frac{dM_d}{dt}\right|_{a_R}.
\end{equation}

\section{Conservation of angular momentum during moonlet spawning}{\label{appendix_spawning}}
The angular momentum of the inner disk before fragmentation is given by equation (\ref{equ_InnerDiskAngularMomentum}). After fragmentation, it becomes
\begin{equation}
 L_d' = \frac{4}{5}M_d'\sqrt{GM_\Earth}\frac{r_d'^{5/2}-R^{5/2}}{r_d'^2-R^2},
\end{equation}
where $M_d'=M_d-m_f$. The angular momentum of the fragment is
\begin{equation}
 L_f=m_f\sqrt{a_fGM_\Earth\left(1-e_f^2\right)},
\end{equation}
where $a_f$ and $e_f$ are its semi-major axis and eccentricity. We set the latter to the ratio of the fragment's escape velocity to the local orbital velocity \citep{lissauer93}
\begin{equation}
 e_f=\frac{\sqrt{2Gm_f/R_f}}{a_f\sqrt{GM_\Earth/a_f^3}}=\sqrt{\frac{2m_fa_f}{M_\Earth R_f}},
\end{equation}
where $R_f$ is the fragment's radius. The fragment's angular momentum then reads
\begin{equation}
 L_f=m_f\sqrt{\frac{a_f}{R_f}GM_\Earth\left( R_f-2\frac{m_f}{M_\Earth}a_f\right)}.
\end{equation}
To compute the new disk's outer edge, we numerically solve for $L_d'+L_f-L_d=0$ so that angular momentum is conserved to a $10^{-8}$ precision. We then move the new body around its orbit so that its actual distance to the new disk's outer edge slightly exceeds its physical radius. Finally, a spawned moonlet's initial inclination is set to half its eccentricity.

\section{Disk-satellite interactions}{\label{appendix_resonances}}
The torque on an exterior moon due to the $(m:m-1)$ inner Lindblad resonance is \citep{goldreich80}
\begin{equation}\label{equ_torque_lindblad}
\Gamma_m = \frac{\pi^2\sigma}{3\Omega\Omega_{ps}}\Psi^2 = \frac{\pi^2\sigma}{3\Omega\Omega_{ps}}\left[ r\frac{d\Phi_m}{dr} + \frac{2\Omega}{\Omega-\Omega_{ps}}\Phi_m\right]^2,
\end{equation}
where $\Omega$ is the orbital frequency in the disk at distance $r$, $\Omega_{ps}$ is the pattern speed of the resonance, and $\Phi_m$ is the $m^{th}$-order Fourier component of the satellite's potential. For $0^{th}$ order inner Lindblad resonances, $\Omega_{ps} = \Omega_s$ where $\Omega_s$ is the satellite's orbital frequency, and the satellite potential can be expressed as \citep{goldreich78}
\begin{equation}
\Phi_m = -\frac{GM_s}{a_s}b_{1/2}^{(m)}(\alpha),
\end{equation}
where $M_s$ and $a_s$ are the satellite's mass and semi-major axis, $\alpha = r/a_s = \left(1-1/m\right)^{2/3}$ and $b_{1/2}^{(m)}(\alpha)$ is the Laplace coefficient of order $1/2$ defined by
\begin{equation}
b_s^{(m)}(\alpha) = \frac{2}{\pi}\int_0^\pi\frac{\cos(m\theta)d\theta}{(1-2\alpha\cos\theta+\alpha^2)^s}.
\end{equation}
With $\Omega = m\Omega_s/(m-1)$, the torque can then be expressed as
\begin{equation}
\Gamma_m = \frac{\pi^2\sigma}{3\Omega_s^2} \left(\frac{m-1}{m}\right) \left(\frac{GM_s}{a_s}\right)^2
           \left(\alpha\frac{db_{1/2}^{(m)}}{d\alpha} + 2mb_{1/2}^{(m)}\right)^2.
\end{equation}
The torque per unit of satellite mass is
\begin{equation}
\frac{\Gamma_m}{M_s} = \frac{\pi^2}{3}\mu_sG\sigma a_sc_m,
\end{equation}
with $\mu_s = M_s/M_{\Earth}$ and $c_m = \alpha^{3/2}\left(\alpha\frac{db_{1/2}^{(m)}}{d\alpha} + 2mb_{1/2}^{(m)}\right)^2$. We then use the following approximation \citep{goldreich80}
\begin{equation}
\left(\alpha\frac{db_{1/2}^{(m)}}{d\alpha} + 2mb_{1/2}^{(m)}\right) 
\approx \frac{2m}{\pi}\left[ K_1\left(\frac{2}{3}\right) + 2K_0\left(\frac{2}{3}\right)\right] 
\approx \frac{2m}{\pi} 2.51,
\end{equation}
where $K_0$ and $K_1$ are modified Bessel functions, so that $c_m \approx 2.55 m^2\left(1-1/m\right)$.

The total torque $T_s$ exerted by the inner disk on an exterior satellite per unit satellite mass is found by summing the torques due to all the $0^{th}$ order resonances that fall in the disk,

\begin{equation}
\frac{T_s}{M_s} = \frac{\displaystyle{\sum_{m=2}^{m_*}} \Gamma_m}{M_s} = \left(\frac{\pi^2}{3}\mu_sG\sigma a_s\right)C(m),
\label{ap_equ_disktorque_on_sat}
\end{equation}
where
\begin{equation}
C(m) = \sum_{m=2}^{m_*}c_m
\end{equation}
and
\begin{equation}
m_* = \left\lfloor\left(1-\left(\frac{r_d}{a_s}\right)^{3/2}\right)^{-1}\right\rfloor,
\end{equation}
where $\lfloor X\rfloor$ is the largest integer not greater than $X$.

The total torque on the disk due to $N$ orbiting moonlets is $\displaystyle{T_d = {- \sum_{s=1}^{N}T_s}}$. For an inner disk with a uniform surface density, changing the disk's angular momentum must involve a change in its outer edge $r_d$ and/or mass flow across its inner boundary.  Because resonances with the outer moonlets generally occur in the outer regions of the disk, we assume that moonlet torques cause a change in the disk's outer edge $r_d$, with $\dot{r}_d|_{moon} < 0$ because external moons remove angular momentum from the disk, with
\begin{equation}
T_d = \frac{dL_d}{dt} = \frac{2}{5}M_d\sqrt{GM_{\Earth}}\left[ \frac{5r_d^{3/2}\dot{r}_d}{r_d^2 - R^2} 
                                                                     - \frac{r_d^{5/2} - R^{5/2}}{\left( r_d^2 - R^2\right)^2}4r_d\dot{r}_d
                                                                \right].
\end{equation}
The rate of change of the disk's outer edge due to $T_d$ is then
\begin{align}
\nonumber\left. \dot{r}_d\right|_{moon} & = \frac{5T_d}{2M_d}\sqrt{\frac{R}{G M_{\Earth}}}
                                 \left[\frac{\left(x^2-1\right)^2}{5x^{3/2}\left(x^2-1\right) - \left(x^{5/2}-1\right)4x}\right]\\
& = \frac{5T_d}{2M_d}\sqrt{\frac{R}{G M_{\Earth}}}g(x).
\label{rdotmoon}
\end{align}

\section{Tidal accretion criteria}{\label{appendix_tidal_accretion}}
At each time step, we detect which particles are about to collide by checking if their relative distance will get smaller than the sum of their radii within the next time step. For each pair of colliding particles, we compute the Jacobi energy after the collision \citep{canup95}:
\begin{equation}
 E_J = \frac{1}{2}\epsilon^2 v_{imp}^2 - \frac{3}{2}x_p^2+\frac{1}{2}z_p^2-\frac{3}{r_p}+\frac{9}{2},
\label{equ_jacobi_energy}
\end{equation}
where $x_p$, $y_p$ and $z_p$ are the Hill coordinates of the impact point, and $v_{imp}$ is the relative impact velocity in units of the Hill velocity $R_H\Omega$, where $\Omega=\sqrt{GM_{\Earth}/a_0^3}$ and
\begin{equation}
 R_H = a_0\left( \frac{m_1+m_2}{3M_{\Earth}} \right)^{1/3}.
\end{equation}
$a_0$ is the local reference radius, $m_1$ and $m_2$ are the masses of the colliding particles with physical radii $r_1$ and $r_2$, $r_p = (r_1+r_2)/R_H$, and $\epsilon$ is an effective coefficient of restitution given by
\begin{equation}
\epsilon = \sqrt{\frac{\epsilon_n^2 v_n^2 + \epsilon_t^2 v_t^2}{v_{imp}^2}},
\end{equation}
where $\epsilon_n$ and $\epsilon_t$ are the normal and tangential coefficients of restitution, and $v_n$ and $v_t$ are the normal and tangential components of $v_{imp}$. If the post-impact Jacobi energy is $< 0$ we assume the collision will result in a perfect merger \citep{ohtsuki93,canup95}. 

\subsection{Angle-averaged criterion}
When averaging equation (\ref{equ_jacobi_energy}) over all possible impact orientations (radial, vertical, and azimuthal), the post-impact Jacobi energy becomes \citep{canup95}
\begin{equation}
E_J = \frac{1}{2}\epsilon^2 v_{imp}^2 - \frac{3}{r_p}-\frac{1}{3}r_p^2+\frac{9}{2}.
\label{equ_Jacobi_Energy_averaged}
\end{equation}
In the limit that $\epsilon=0$ (a completely inelastic collision), requiring $E_J < 0$ for accretion yields $r_p < 0.7$.

\subsection{Total accretion criterion}
An alternative is to assume that the particles are aligned in the radial direction, which is the widest dimension of the Hill ``sphere'' and therefore the most favorable for growth.  In this case, equation (\ref{equ_jacobi_energy}) becomes
\begin{equation}
E_J = \frac{1}{2}\epsilon^2 v_{imp}^2 - \frac{3}{r_p}-\frac{3}{2}r_p^2+\frac{9}{2}.
\label{equ_Jacobi_Energy_total}
\end{equation}
In the limit that $\epsilon=0$, requiring $E_J < 0$ for accretion then yields $r_p < 1$.

\subsection{Minimal distance for accretion}
\begin{figure}[!h]
\begin{center}
\includegraphics[width=8cm]{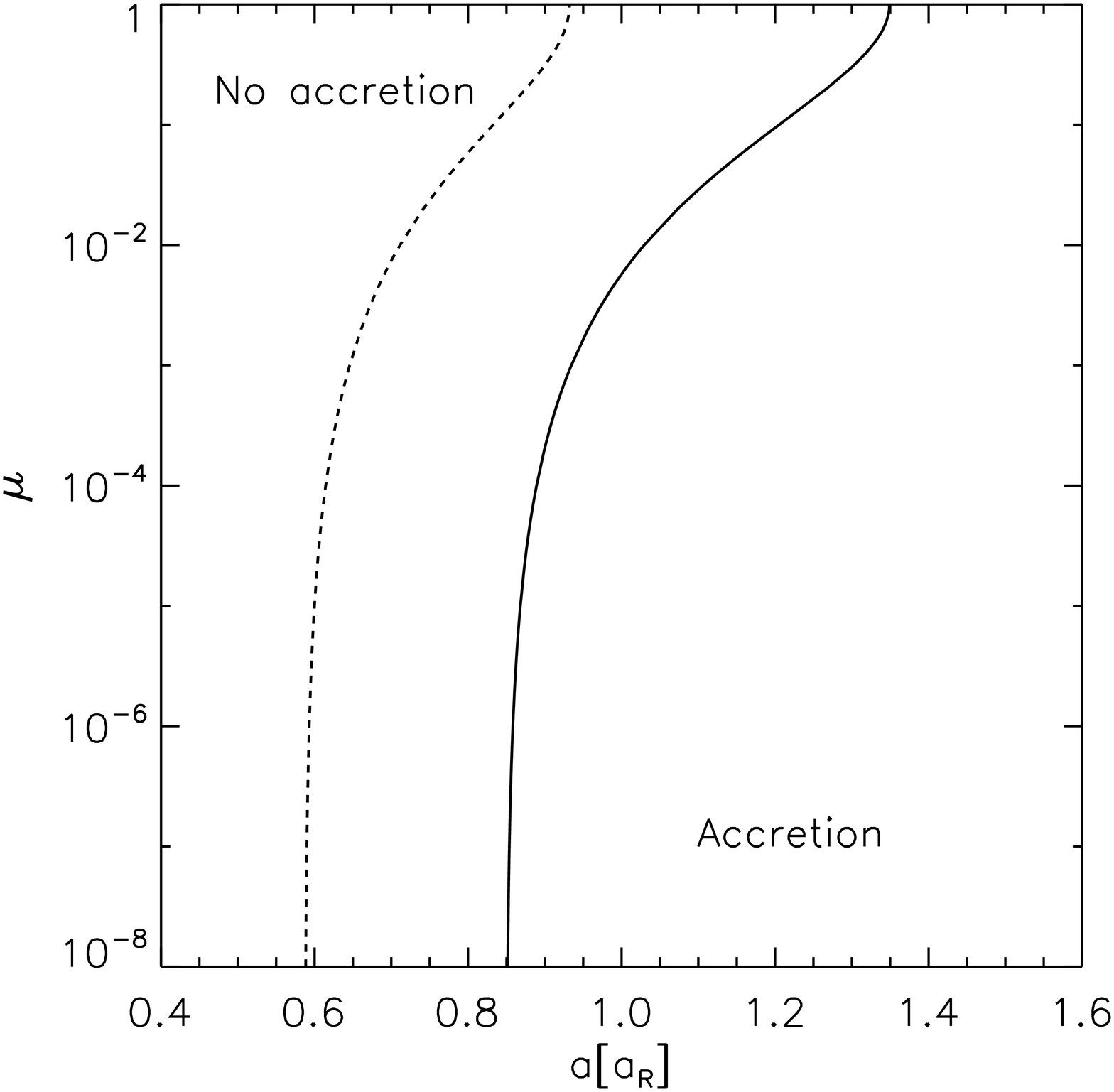}
\end{center}
\caption{Orbital distance (in units of the Roche radius, $a_R$) beyond which two colliding particles with mass ratio $\mu$ may accrete. The model assumes an inelastic collision between two spherical particles. The solid line corresponds to an average over all possible impact orientations, while the dashed line considers a purely radial collision \citep{canup95}.}
\label{figAccretionCondition}
\end{figure}

The quantity $r_p=(r_1+r_2)/R_H$ can be expressed as 
\begin{equation}
r_p = \frac{R_c}{a_0}\left( \frac{\rho}{3\rho_c} \right)^{-1/3} \frac{1 + \mu^{1/3}}{\left(1 + \mu \right)^{1/3}}
\approx 0.6 \frac{a_R}{a_0} \frac{1 + \mu^{1/3}}{\left(1 + \mu \right)^{1/3}},
\end{equation}
where $R_c$ and $\rho_c$ are the radius and bulk density of the central body, $\rho$ is the material density of the colliding particles, $a_R=2.456R_c(\rho_c/\rho)^{1/3}$ is the Roche limit for material density $\rho$, and $0 < \mu \leq 1$ is the mass ratio of colliding particles. Using the above constraints on $r_p$, we can derive a minimum distance beyond which two particles of a given mass ratio $\mu$ can accrete, depending on $v_{imp}$ and $\epsilon$. This is represented in Figure \ref{figAccretionCondition}.

If the post-impact Jacobi energy of the colliding particles is positive, we assume they rebound inelastically and do not merge.  The relative velocity of the particles is then modified as
\begin{equation}
\begin{cases}
v_n' = -\epsilon_n v_n\\
v_t' = \epsilon_t v_t
\end{cases}
\end{equation}
where $v_n'$ and $v_t'$ are the post-impact normal and tangential velocity components.


\newpage


\begin{thebibliography}{}
 \bibitem[Benz et al.(1989)]{benz89}
Benz, W., Cameron, A.~G.~W., \& Melosh, H.~J. 1989, \icarus, 81, 113

\bibitem[Benz et al.(1986)]{benz86}
Benz, W., Slattery, W.~L., \& Cameron, A.~G.~W. 1986, \icarus, 66, 515

\bibitem[Benz et al.(1987)]{benz87}
Benz, W., Slattery, W.~L., \& Cameron, A.~G.~W. 1987, \icarus, 71, 30

\bibitem[Cameron(1997)]{cameron97}
Cameron, A.~G.~W. 1997, \icarus, 126, 126

\bibitem[Cameron \& Benz(1991)]{cameron91}
Cameron, A.~G.~W., \& Benz, W. 1991, \icarus, 92, 204

\bibitem[Cameron \& Ward(1976)]{cameron76}
Cameron, A.~G.~W., \& Ward, W.~R. 1976, Lunar Sci. Abst., VII, 120

\bibitem[Canup(2004)]{canup04a}
Canup, R.~M. 2004a, \araa, 42, 441

\bibitem[Canup(2004)]{canup04b}
Canup, R.~M. 2004b, \icarus, 168, 433

\bibitem[Canup(2008)]{canup08}
Canup, R.~M. 2008, \icarus, 196, 518

\bibitem[Canup \& Asphaug(2001)]{canup01}
Canup, R.~M., \& Asphaug, E. 2001, \nat, 412, 708

\bibitem[Canup et al.(2012)]{canup12}
Canup, R.~M., Barr, A.~C., \& Crawford, D.~A. 2012, Submitted to \icarus

\bibitem[Canup \& Esposito(1995)]{canup95}
Canup, R.~M., \& Esposito, L.~W. 1995, \icarus, 113, 331

\bibitem[Canup et al.(1999)]{canup99}
Canup, R.~M., Levison, H.~F., \& Stewart, G.~R. 1999, \aj, 117, 603

\bibitem[Canup \& Ward(2000)]{canup00}
Canup, R.~M., \& Ward, W.~R. 2000, Proc. Lunar planet  Sci. XXXI, Abst. 1916

\bibitem[Chambers(2001)]{chambers01}
Chambers, J.~E. 2001, \icarus, 152, 205

\bibitem[Charnoz et al.(2010)]{charnoz10}
Charnoz, S., Salmon, J., \& Crida, A. 2010, \nat, 465, 752

\bibitem[Clayton(1993)]{clayton93}
Clayton, R.~N. 1993, Annu. Rev. Earth Planet. Sci., 21, 115

\bibitem[Daisaka \& Ida(1999)]{daisaka99}
Daisaka, H., \& Ida, S. 1999, Earth, Planets, and Space, 51, 1195

\bibitem[Daisaka et al.(2001)]{daisaka01}
Daisaka, H., Tanaka, H., \& Ida, S. 2001, \icarus, 154, 296

\bibitem[Dermott et al.(1988)]{dermott88}
Dermott, S.~F., Malhotra, R., \& Murray, C.~D. 1988, \icarus, 76, 295

\bibitem[Duncan et al.(1998)]{duncan98}
Duncan, M.~J., Levison, H.~F., \& Lee, M.~H. 1998, \aj, 116, 2067

%\bibitem[Fernandez \& Ip(1984)]{fernandez84}
%Fernandez, J.~A., \& Ip, W.~H. 1984, \icarus, 58, 109

\bibitem[Goldreich \& Tremaine(1980)]{goldreich80}
Goldreich, P., \& Tremaine, S. 1980, \apj, 241, 425

\bibitem[Goldreich \& Tremaine(1978)]{goldreich78}
Goldreich, P., \& Tremaine, S.~D. 1978, \icarus, 34, 240

\bibitem[Goldreich \& Ward(1973)]{goldreich73}
Goldreich, P., \& Ward, W.~R. 1973, \apj, 183, 1051

%\bibitem[Hahn \& Malhotra(1999)]{hahn99}
%Hahn, J. M., \& Malhotra, R. 1999, \aj, 117, 3041

\bibitem[Hartmann \& Davis(1975)]{hartmann75}
Hartmann, W.~K., \& Davis, D.~R. 1975, \icarus, 24, 504

\bibitem[Ida et al.(1997)]{ida97}
Ida, S., Canup, R.~M., \& Stewart, G.~R. 1997, \nat, 389, 353

\bibitem[Kokubo et al.(2000)]{kokubo00}
Kokubo, E., Canup, R.~M., \& Ida, S. 2000, in Origin of the Earth and Moon, ed. Canup, R.~M., \& Righter, K. (Tuscon: Univ. arizona Press), 145

\bibitem[Lissauer \& Stewart(1993)]{lissauer93}
Lissauer, J.~J., \& Stewart, G.~R. 1993, in Planets Around Pulsars, ed. Phillips, J.~A., Thorsett, S.~E. \& Kulkarni, S.~R., 217

\bibitem[Ohtsuki(1993)]{ohtsuki93}
Ohtsuki, K. 1993, \icarus 106, 228

\bibitem[Pahlevan \& Stevenson(2007)]{pahlevan07}
Pahlevan, K., \& Stevenson, D.~J. 2007, Earth Planet. Sci. Lett., 262, 438

\bibitem[Papaloizou \& Larwood(2000)]{papaloizou00}
Papaloizou, J.~C.~B., \& Larwood, J.~D. 2000, \mnras, 315, 823

\bibitem[Papaloizou et al. (2007)]{papaloizou07}
Papaloizou, J.~C.~B., Nelson, R.~P., Kley, W., Masset, F.~S., \& Artymowicz, P. 2007, in Protostars and Planets V, ed. Reipurth, B., Jewitt, D., \& Keil, k. (Tuscon: Univ. arizona Press).

\bibitem[Pritchard \& Stevenson(2000)]{pritchard00}
Pritchard, M.~E., \& Stevenson, D.~J. 2000, in Origin of the Earth and Moon, ed. Canup, R.~M., \& Righter, K. (Tuscon: Univ. arizona Press), 179

\bibitem[Salmon et al.(2010)]{salmon10}
Salmon, J., Charnoz, S., Crida, A., \& Brahic, A. 2010, \icarus, 209, 771

\bibitem[Salo(1995)]{salo95}
Salo, H. 1995, \icarus, 117, 287

\bibitem[Sridhar \& Tremaine(1992)]{sridhar92}
Sridhar, S., \& Tremaine, S. 1992, \icarus, 95, 86

\bibitem[Stevenson(1987)]{stevenson87}
Stevenson, D.~J. 1987, Annu. Rev. Earth Planet. Sci., 15, 271

\bibitem[Takeda \& Ida(2001)]{takeda01}
Takeda, T., \& Ida, S. 2001, \apj, 560, 514

\bibitem[Thompson \& Stevenson(1988)]{thompson88}
Thompson, C., \& Stevenson, D.~J. 1988, \apj, 333, 452

\bibitem[Toomre(1964)]{toomre64}
Toomre, A. 1964, \apj, 139, 1217

\bibitem[Trinquier  et al.(2009)]{trinquier09}
Trinquier, A., Elliott, T., Ulfbeck, D., Coath, C., Krot, A.~N., \& Bizzarro, M. 2009, Science 324, 374

%\bibitem[Walsh \& Morbidelli(2011)]{walsh11}
%Walsh, K.~J., \& Morbidelli, A. 2011, \aap, 526, A126

\bibitem[Ward(2012)]{ward12}
Ward, W.~R. 2012, \apj, 744, 140

\bibitem[Ward \& Cameron(1978)]{ward78}
Ward, W.~R., \& Cameron, A.~G.~W. 1978, Proc. Lunar planet  Sci. IX, Abst. 1205

\bibitem[Ward \& Canup(2000)]{ward00}
Ward, W.~R., \& Canup, R.~M. 2000, \nat, 403, 741

\bibitem[Wiechert et al.(2001)]{wiechert01}
Wiechert, U., Halliday, A.~N., Lee, D.~C., Snyder, G.~A., Taylor, L.~A., \& Rumble, D. 2001, Science, 294, 345

\bibitem[Zhang et al.(2012)]{zhang12}
Zhang, J., Dauphas, N., Davis, A.~M., Leya, I., \& Fedkin, A. 2012, Nature Geosci., 5, 224
\end{thebibliography}
\end{document}